\documentclass[final,notitlepage,12pt,letterpaper]{article}
\usepackage{makeidx}
\usepackage{amssymb}
\usepackage{amsmath}
\usepackage[dvips]{graphicx}
\usepackage{amsmath}
\usepackage{epsfig}
\usepackage[dvips]{geometry}
\usepackage{setspace}
\usepackage{endnotes}

\input{tcilatex}
\begin{document}

\title{Fiscal Stimulus of Last Resort\vspace{-0.15in}}
\author{Alessandro Piergallini\thanks{%
Associate Professor of Economics, Department of Economics and Finance,
University of Rome Tor Vergata, Via Columbia 2, 00133 Rome, Italy. ORCID:
0000-0001-6458-0364. Phone: +390672595639. E-mail:
alessandro.piergallini@uniroma2.it. Homepage: https://sites.google.com/
view/alessandropiergallini. I wish to thank Paolo Canofari, Alessia
Franzini, Giovanni Piersanti, Michele Postigliola and Luca Vitali for very
useful comments. The usual disclaimers apply.}\vspace{-0.1in} \\
%EndAName
\textit{University of Rome Tor Vergata}}
\maketitle

\vspace{-0.25in}

\begin{abstract}
I examine global dynamics in a monetary model with overlapping generations
of finite-horizon agents and a binding lower bound on nominal interest
rates. Debt targeting rules exacerbate the possibility of self-fulfilling
liquidity traps, for agents expect austerity following deflationary slumps.
Conversely, activist but sustainable fiscal policy regimes---implementing
intertemporally balanced tax cuts and/or transfer increases in response to
disinflationary trajectories---are capable of escaping liquidity traps and
embarking inflation into a globally stable path that converges to the
target. Should fiscal stimulus of last resort be overly aggressive, however,
spiral dynamics around the liquidity-trap steady state exist, causing global
indeterminacy.

\vspace{0.1in}

\noindent \textbf{JEL Classification:} E31; E62; E63.

\noindent \textbf{Keywords}: Monetary-Fiscal Policy Interactions;
Overlapping Generations; Intergenerational Wealth Effects; Multiple
Equilibria; Global Dynamics.
\end{abstract}

\newpage

%\begin{doublespace}

\section{Introduction}

The recent observed plunge in global monetary policy rates towards zero or
below both as a result of the secular stagnation (Summers, 2016, 2018;
Krugman, 2020) and in the attempt to cope with the corona crisis---in a
macroeconomic environment featured by effective and expected inflation rates
systematically undershooting their targets for an extended period in most
major economies following the Great Recession (Bartsch et al., 2019)---poses
salient challenges for public policy design. A central question that
arguably urges for macroeconomic analysis is how to escape unintended
deflationary slumps associated to the existence of a liquidity-trap
equilibrium when the economy lies in the vicinity of the effective lower
bound policy rate, whereby conventional monetary policy reveals to be
impotent in fostering aggregate demand and therefore the level of prices,
without triggering other sources of macroeconomic instability---such as
unsustainability of fiscal policy or a burst of inflation. The contribution
of the present paper is to analyze the issue of aggregate stability in an
overlapping generations monetary model that exhibits multiple steady-state
equilibria because of the existence of a binding lower bound on nominal
interest rates. By departing from the Ricardian debt equivalence (Barro,
1974), due to a finite planning horizon of private agents and to the absence
of perfect intergenerational altruism as originally proposed in the Yaari
(1965) and Blanchard (1985)'s uncertain lifetime approach, the setup
developed in this study enables me to explore in an analytically tractable
way the effects of alternative fiscal policy regimes and their interaction
with monetary policy from both a local- and a global-dynamics perspective.

The paper establishes policy-relevant results that would not appear in the
traditional infinitely-lived single representative agent framework. First I
show the dynamic consequences of a budgetary policy regime in which the
fiscal authority gradually adjusts the stock of real government liabilities
relative to the size of the economy in order to converge towards a target
level in the long run, as assumed by Minea and Villieu (2013), Maebayashi,
Hori and Futagami (2017), and Cheron et al. (2019) in non-monetary contexts
with endogenous growth to analyze the case of fiscal constraints in the
spirit of the Maastricht-Treaty framework prevailing in the European Union.
I demonstrate that such a regime of stringent fiscal discipline makes the
economic system prone to self-fulfilling deflationary (or disinflationary)
trajectories approaching to an unintended steady state that exhibits the
features of a liquidity trap, which prevents the monetary authority from
uniquely determining inflation along a saddle path passing through the
target rate.

In the present setup with dyachronous heterogeneity, deflationary dynamics
induce, \textit{per se}, a redistribution of real financial wealth from
future to current generations, because the implied increase in the real
value of government liabilities rises the burden of future fiscal
restrictions. Such an intergenerational redistribution of wealth in favor of
currently alive generations would provide, \textit{ceteris paribus}, a
stimulus to aggregate demand and thus to prices, tending as a consequence to
counter-react to the initial deflationary pressures. In other words, a
stabilizing role of wealth effects, potentially capable of making liquidity
traps implausible along the lines of the traditional theory of Pigou (1943,
1947) and Patinkin (1965), is operative in my optimizing framework.

However, in a condition of general equilibrium that takes also into account
the design of fiscal policy, if the government is engaged in making the
stock of real government liabilities relative to the size of the economy
converge to a target level to comply constraints of the Maastricht-type,
following deflation primary surpluses must increase in order to avoid
escalation of real debt. Under these circumstances, the resulting fiscal
austerity measures generate a negative wealth effect on current cohorts'
consumption and turn to reinforce the initial deflationary pressures, hence
leading the economic system to a liquidity-trap steady state, whereby
monetary policy loses control over the inflation rate. Global indeterminacy
arises. Specifically, any inflation trajectory originating below the saddle
path passing through the target steady state can be sustained as an
equilibrium outcome, bringing about macroeconomic instability.

Then I explore the issue of how to lift the economy out of liquidity traps.
I demonstrate that a large class of `activist' but sustainable fiscal policy
regimes---decreasing taxes and/or increasing public transfers in response to
decelerating inflation under the respect of the government's solvency
condition---enable the economic system to escape liquidity traps and, at the
same time, embark inflation into a globally stable path converging to the
target rate. My results indicate that to sustain global determinacy and thus
ensure macroeconomic stability, there is no need for the government to
render the liquidity-trap steady state fiscally unsustainable as predicted
in the context of the traditional Ramsey-type single representative agent
paradigm.

Should deflationary spirals take place, a sequence of fiscal expansions
financed by issuance of bonds and future taxes net of transfers enlarges
both human and financial wealth for currently alive households, thereby
sustaining aggregate consumption and exerting a powerful stimulus to
aggregate demand despite the expected increase in real interest rates in the
vicinity of the trapping equilibrium, for government debt is net wealth for
living generations. This course of sustainable policy action, which partly
shifts the sequence of future net taxes to future generations thereby
inducing current generations to dissave, is shown to be capable of
offsetting self-fulfilling falls in prices, restoring in general equilibrium
convergence of inflation towards the intended steady state.

However, I demonstrate that not all types of expansionary budgetary policies
that are able to eradicate liquidity traps are compatible with global
determinacy. In particular, I find that when fiscal stimulus is overly
aggressive, (unstable) spiral dynamics around the liquidity-trap steady
state exist. In this case, even though convergence of inflation towards the
target rate along a saddle connection is ensured, global indeterminacy
prevails. In addition, the inflation rate may fluctuate for relatively long
periods of time in a region whereby monetary policy is necessarily
`passive'---in the sense of Leeper (1991)---because of the effective lower
bound on the nominal interest rate, away from the intended steady state.
These results cast doubts on the stabilizing properties of excessively lax
fiscal boosts expected to be backed by large future net taxes required to
redeem the government debt eventually.

Two clear-cut policy implications stem from my analytical findings. First,
strict fiscal discipline does not support price stability, contrary to
conventional wisdom, at least since Sargent and Wallace (1981). Rather, it
is a potential source of macroeconomic imbalances, for it dampens aggregate
demand following deflationary slumps, thus exacerbating expectations-driven
instabilities in prices. A self-fulfilling `austerity-deflation' nexus does
amplify the possibility of undesired liquidity traps. Second, enforcing an
intertemporally balanced and moderately aggressive bond-financed fiscal
stimulus at a time of out-of-equilibrium deflationary trajectories plays an
essential role, complementary to inflation-targeting-oriented interest rate
feedback rules (e.g., Taylor, 1999, 2012; Woodford, 2003; {Gal\'{\i},} 2015;
Walsh, 2017), in order to guarantee price and macroeconomic stability.
Central banks can determine equilibrium inflation globally under the
rules-based approach to monetary policy (Taylor, 2021), but only if
supported by feedback budgetary policy actions of `last resort'.

The paper is organized as follows. Section 2 points out the paper's
connections with the literature and elucidates the novel contribution of the
present study. Section 3 develops the optimizing continuous time overlapping
generations model and specifies the monetary policy regime. Section 4
examines global dynamics under a debt targeting fiscal policy regime.
Section 5 examines global dynamics under an activist fiscal policy regime
compatible with the respect of the government's intertemporal budget
constraint. The concluding Section 6 sums up the main results.

\section{Related Literature}

The macroeconomic framework set forth in this paper and the results that
emerge from the present analysis are markedly different than those
prevailing in most of the literature on `confidence-driven' liquidity traps
(see Bilbie, 2018, and Nakata and Schmidt, 2019). The seminal works by
Benhabib, Schmitt-Groh\'{e} and Uribe (2001, 2002) and Schmitt-Groh\'{e} and
Uribe (2009) employ an infinitely-lived single representative agent setup
with real balance effects \`{a} la Sidrauski (1967)-Brock (1974, 1975),
thereby overlooking intergenerational heterogeneity and the associated
wealth effects on aggregate demand dynamics. The analysis developed in this
paper is an effort to fill this gap. In particular, Benhabib, Schmitt-Groh%
\'{e} and Uribe (2002) show that fiscal policy requires to be unsustainable
at the steady state exhibiting deflation or relatively low inflation in
order to avoid the liquidity trap equilibrium, for the agents'
transversality condition turns to be violated. In contrast, in the present
setting with overlapping generations of finitely-lived agents, which brings
the advantage of encompassing the Ramsey-type framework as a limiting case,
I demonstrate that fiscal boosts satisfying both the individuals'
transversality condition and the government's solvency constraint do suffice
to escape liquidity traps and, at the same time, sustain global determinacy.
Stabilizing budgetary policies need not be intertemporally unbalanced. In
addition, however, I prove that fiscal stimulus of last resort should not be
overly aggressive, in order to avoid that the liquidity-trap steady state is
an unstable spiral point---which would give rise to global indeterminacy.

Performing a thorough numerical analysis of a nonlinear New Keynesian model
with distortionary taxation, Mertens and Ravn (2014) show that supply-side
policies such as cuts marginal labour tax rates---as opposed of demand-side
policies such as increases in government spending---are necessary to
properly offset expectations-driven liquidity traps. My paper differs in
three relevant dimensions. First, I present an alternative way to analyze
the implications of confidence-driven liquidity traps for fiscal policy
design, since I employ an overlapping generations model whereby demographic
heterogeneity matters---contrary to the standard New Keynesian setup.
Taxes/transfers are lump sum, prices are flexible and time is continuous, in
order to make my analysis directly comparable to Benhabib, Schmitt-Groh\'{e}
and Uribe (2002) and reconsider transparently the role of demand stimulating
budgetary policies in a model where private agents have finite horizons,
without unnecessary complications. Combining distortionary taxes and sticky
prices with an overlapping generations setting and analyzing the implied
global dynamics are important considerations for future research. The
present modeling approach to monetary-fiscal interactions in the presence of
multiple steady-state equilibria could then constitute a fruitful benchmark
for more complex investigations along these lines. Second, in my simple and
tractable continuous-time model I am enabled to explore analytically the
scope for the existence---under a particular type of fiscal policy---of a
saddle connection among different steady states, leading to global
determinacy under the Taylor-rule monetary framework and escaping the
liquidity-trap equilibrium. Third, differently from Mertens and Ravn (2014),
I do find a positive role of demand-side-oriented fiscal stimulus respecting
the government's intertemporal budget constraint in order to restore an
inflationary path uniquely converging to the central bank's target rate. The
view that fiscal intervention stimulating aggregate demand, financed by bond
issues and future higher taxes and/or lower expenditures to avoid public
insolvency, is a powerful tool to reverse downward pressure on inflation
linked to the existence of a liquidity-trap equilibrium appears to have
sound microfoundations.

Using a standard New Keynesian model \`{a} la Woodford (2003), Schmidt
(2016) shows that Ricardian government spending rules that prevent a decline
in real marginal costs following a confidence shock protect the economy from
falling into expectations-driven liquidity traps. My paper is complementary
in three important dimensions. First, my contribution is to demonstrate that
sustainable fiscal actions can be justified even in the most innocuous
monetary framework with flexible prices, once intergenerational
heterogeneity is accounted for. The stabilizing dynamic effect of Ricardian
budgetary policies does not require pricing frictions. Second, in the
present model with overlapping-generations demographics I show that
out-of-target deflationary slumps are typically escaped not by making the
liquidity-trap steady state incompatible with agents' optimizing conditions
in general equilibrium, as in Schmidt (2016), but by making it locally
unstable. Under these circumstances, I prove that there exists a
heteroclinic orbit connecting the unintended steady state with the intended
one, along which global determinacy of equilibrium applies should fiscal
policy design ensure that the undesired steady state is an unstable node.
Third, I analyze the dynamic implications of a Ricardian experiment
concerning an intertemporal reallocation of taxes and transfers, in order to
elucidate the consequences of intergenerational wealth effects in an
otherwise conventional model exhibiting the presence of Ricardian
equivalence and the existence of liquidity traps. Importantly, from an
empirical perspective, such a policy focus is justifiable from the fact that
fiscal stimulus packages in the aftermath of both the Great Recession and
the corona crisis have largely consisted of increases in public transfers
(e.g., Taylor, 2018; Bayer et al., 2020). Remarkably, the first fiscal plan
decided by the Biden administration in the United States---The American
Rescue Plan Act of 2021\footnote{%
https://www.congress.gov/bill/117th-congress/house-bill/1319/text.}%
---consists for approximately three quarters of transfers.

\section{The Model}

For the objectives of this work, I model intergenerational heterogeneity
within a monetary economy in continuous time by setting forth a modified
version of the Yaari (1965)-Blanchard (1985)-Weil (1989) overlapping
generations framework, whereby forward looking agents have finite horizons,
extended to incorporate money in the agents' asset menu and embed the more
general case of non-separable preferences over consumption and real cash
balances. The resulting framework will permit me to remain close to the
literature on global perspectives of macroeconomic policy rules that
originates at least since the seminal contributions by Benhabib, Schmitt-Groh%
\'{e} and Uribe (2001, 2002), based on the other hand upon the
infinite-horizon single representative agent setup.

\subsection{The Individual's Consumption Behavior}

Agents face uncertainty about the duration of their lives. Each individual,
in particular, is assumed to be subject to a common and constant
instantaneous probability of death, $\mu >0$. At each instant of time $t$ a
new generation is born and total population is assumed to grow at a constant
rate $n$. Therefore, the birth rate is $\beta =n+\mu $. Denote by $N(t)$
total population at time $t$, with $N\left( 0\right) =1$ for simplicity.
Hence, the size of the generation born at time $t$\ is $\beta N(t)=\beta
e^{nt}$, while the size of the surviving cohort born at time $s\leq t$\ is $%
\beta N\left( s\right) e^{-\mu (t-s)}=\beta e^{-\mu t}e^{\beta s}$.\
Population at time $t$\ is consequently given by $N(t)=\beta e^{-\mu
t}\int_{-\infty }^{t}e^{\beta s}ds$.

There is no intergenerational operative bequest motive, following Blanchard
(1985). Newly born agents have no assets. Individuals have identical
preferences and are assumed to supply one unit of labor inelastically, which
is transformed one-for-one into output, for analytical convenience. Each
agent belonging to the generation born at time $s\leq 0$ chooses the time
path of consumption, $\overline{c}(s,t)$, and real money balances, $%
\overline{m}(s,t)$, in order to maximize the expected discounted value of
the utility function given by%
\begin{equation}
E_{0}\int_{0}^{\infty }\log \Upsilon \left( \overline{c}(s,t),\overline{m}%
(s,t)\right) e^{-\rho t}dt,  \label{0}
\end{equation}%
where $E_{0}$ denotes the expectation operator conditional on period $0$
information, $\rho >0$ is the pure rate of time preference, and the
subutility function\textit{\ }$\Upsilon \left( \cdot \right) $\textit{\ }is
strictly increasing, concave,\textit{\ }and linearly homogenous.\textit{\ }%
According to Reis (2007), consumption and real money balances are Edgeworth
complements, that is, $\Upsilon _{cm}>0$. Following Cushing (1999), the
elasticity of substitution between the two is lower than unity.

Using the fact that the probability at time $0$ of surviving at time $t\geq
0 $ is $e^{-\mu t}$, the expected lifetime utility function (\ref{0}) can be
rewritten as%
\begin{equation}
\int_{0}^{\infty }\log \Upsilon \left( \overline{c}(s,t),\overline{m}%
(s,t)\right) e^{-\left( \mu +\rho \right) t}dt,  \label{1}
\end{equation}%
implying that the effective subjective discount rate with lifetime
uncertainty is $\mu +\rho $.

Individuals accumulate their real assets, $\overline{a}(s,t)$, in the form
of interest bearing public bonds, $\overline{b}(s,t)$, and real money
balances, so that $\overline{a}(s,t)=\overline{b}(s,t)+\overline{m}(s,t)$.
The instantaneous budget constraint in real terms takes the form%
\begin{equation}
\dot{\overline{a}}(s,t)=\left( R(t)-\pi (t)+\mu \right) \overline{a}(s,t)+%
\overline{y}(s,t)-\overline{\tau }(s,t)-\overline{c}(s,t)-R(t)\overline{m}%
(s,t),  \label{2}
\end{equation}%
where $R(t)$ denotes the nominal interest rate, $\pi (t)$ the inflation
rate, $\overline{y}(s,t)$ output, and $\overline{\tau }(s,t)$ real lump-sum
taxes net of public transfers. Following Yaari (1965), the budget constraint
incorporates the hypothesis that in each period consumers of generation $s$
receive an actuarial fair premium, given by $\mu \overline{a}(s,t)$, from
perfectly competitive life insurance companies in exchange for their total
financial wealth at the time of death. The life insurance market avoids the
possibility for individuals of passing away leaving undesired bequests to
their heirs. As emphasized by Blanchard (1985), under the alternative
hypothesis of actuarial bonds issued by financial intermediaries, results
would be equivalent.

Agents are precluded from engaging in Ponzi's games, implying
\begin{equation}
\underset{t\rightarrow \infty }{\lim }\overline{a}(s,t)e^{-%
\int_{0}^{t}(R(j)-\pi (j)+\mu )dj}\geq 0.  \label{3}
\end{equation}

Let $\overline{x}(s,t)$ denote total consumption, defined as physical
consumption plus the interest forgone on real money holdings. That is,
\begin{equation}
\overline{x}(s,t)\equiv \overline{c}(s,t)+R(t)\overline{m}(s,t).  \label{tc}
\end{equation}%
Hence, the agent's optimizing problem can be solved using a two-stage
budgeting procedure (Deaton and Muellbauer, 1980; Marini and van der Ploeg,
1988).

In the first stage, consumers solve an intratemporal maximizing problem of
choosing the efficient allocation between consumption and real money
balances, in order to maximize the instantaneous subutility function $%
\Upsilon \left( \cdot \right) $ for a given level of total consumption.
Optimality implies that the marginal rate of substitution between
consumption and real money balances must equal the nominal interest rate:
\begin{equation}
\frac{\Upsilon _{m}\left( \overline{c}(s,t),\overline{m}(s,t)\right) }{%
\Upsilon _{c}\left( \overline{c}(s,t),\overline{m}(s,t)\right) }=R(t).
\label{intra}
\end{equation}%
Because preferences are linearly homogenous, condition (\ref{intra}) takes
the form%
\begin{equation}
\overline{c}(s,t)=\Omega \left( R(t)\right) \overline{m}(s,t),  \label{4}
\end{equation}%
where $\Omega ^{\prime }\left( R(t)\right) >0$. This sign restriction
follows from $\Upsilon _{cc}-\Upsilon _{cm}\Upsilon _{c}/\Upsilon _{m}<0$
and $\Upsilon _{mm}-\Upsilon _{cm}\Upsilon _{m}/\Upsilon _{c}<0$.

In the second stage, agents solve an intertemporal maximizing problem of
choosing the time path of total consumption, $\overline{x}(s,t)$, in order
to maximize their lifetime utility function (\ref{1}) given the constraints (%
\ref{2})-(\ref{3}) and the optimal static condition (\ref{4}). At optimum%
\footnote{%
See Appendix A for analytical details.}%
\begin{equation}
\dot{\overline{x}}(s,t)=\left( R(t)-\pi (t)-\rho \right) \overline{x}(s,t),
\label{4a}
\end{equation}%
\begin{equation}
\underset{t\rightarrow \infty }{\lim }\overline{a}(s,t)e^{-%
\int_{0}^{t}(R(j)-\pi (j)+\mu )dj}=0.  \label{4c}
\end{equation}

Integrating forward the instantaneous budget constraint (\ref{2}) after
using definition (\ref{tc}), applying the transversality condition (\ref{4c}%
) and employing the dynamic equation (\ref{4a}), total consumption can be
expressed as a linear function of total wealth:%
\begin{equation}
\overline{x}(s,t)=(\mu +\rho )\left( \overline{a}(s,t)+\overline{h}%
(s,t)\right) ,  \label{4d}
\end{equation}%
where
\begin{equation}
\overline{h}(s,t)\equiv \int_{t}^{\infty }\left( \overline{y}(s,v)-\overline{%
\tau }(s,v)\right) e^{-\int_{t}^{v}(R(j)-\pi (j)+\mu )dj}dv  \label{human}
\end{equation}%
quantifies human wealth, that is, the present discounted value of
after-tax/transfers labor income.

Using (\ref{tc}), (\ref{4}), and (\ref{4d}), it also follows that individual
phisical consumption is a function of total wealth:%
\begin{equation}
\overline{c}(s,t)=\frac{(\mu +\rho )}{\Lambda (R(t))}\left( \overline{a}%
(s,t)+\overline{h}(s,t)\right) .  \label{4e}
\end{equation}%
where $\Lambda \left( R(t)\right) \equiv 1+R(t)/\Omega \left( R(t)\right) $.
Combining (\ref{tc}), (\ref{4}) and (\ref{4a}) yields the optimal time path
of individual consumption:%
\begin{equation}
\dot{\overline{c}}(s,t)=\left[ \left( R(t)-\pi (t)-\rho \right) -\frac{%
\Lambda ^{\prime }(R(t))}{\Lambda (R(t))}\dot{R}(t)\right] \overline{c}(s,t),
\label{5}
\end{equation}%
where $\Lambda ^{\prime }\left( R(t)\right) >0$. This sign restriction
follows from $\Omega \left( R(t)\right) -R(t)\Omega ^{\prime }\left(
R(t)\right) >0$, which in turn depends on the elasticity of substitution
between real money balances and consumption, $\Omega ^{\prime }\left(
R(t)\right) R(t)/\Omega \left( R(t)\right) $, assumed to be lower than
unity. Notice that, according to (\ref{5}), the growth rate of optimal
consumption is identical across all generations, for it is independent of $s$%
.

\subsection{Time Paths of Aggregate Variables}

I can now derive the evolution of aggregate variables. The population
aggregate for a generic variable at individual level $\overline{z}(s,t)$ can
be obtained by integrating over all generations, so that%
\begin{equation}
\overline{Z}(t)\equiv \beta e^{-\mu t}\int_{-\infty }^{t}\overline{z}%
(s,t)e^{\beta s}ds,  \label{aggregation}
\end{equation}%
where the upper case letter indicates the aggregate value at the population
level. The corresponding variable in per capita terms is denoted as%
\begin{equation}
\overline{z}(t)\equiv \overline{Z}(t)e^{-nt}=\beta \int_{-\infty }^{t}%
\overline{z}(s,t)e^{\beta \left( s-t\right) }ds.  \label{per-capita}
\end{equation}

For analytical convenience, assume that each agent faces identical
age-independent income and net tax flows, so that $\overline{y}(s,t)=%
\overline{y}(t)$\ and $\overline{\tau }(s,t)=\overline{\tau }(t)$, as in
Blanchard (1985).\textit{\ }Then, using $\overline{a}(t,t)=0$ and
consequently $\overline{c}(t,t)=\left[ \alpha (\mu +\rho )/\Lambda (R(t))%
\right] \overline{h}(t,t)$, the budget constraint and the optimal time path
of consumption expressed in per capita terms are given by, respectively,%
\footnote{%
See Appendix B for analytical details.}%
\begin{equation}
\dot{\overline{a}}(t)=\left( R(t)-\pi (t)-n\right) \overline{a}(t)+\overline{%
y}(t)-\overline{\tau }(t)-\overline{c}(t)-R(t)\overline{m}(t),  \label{6}
\end{equation}%
\begin{equation}
\dot{\overline{c}}(t)=\left[ \left( R(t)-\pi (t)-\rho \right) -\frac{\Lambda
^{\prime }(R(t))}{\Lambda (R(t))}\dot{R}(t)\right] \overline{c}(t)-\frac{%
\beta (\rho +\mu )}{\Lambda (R(t))}\overline{a}(t),  \label{7}
\end{equation}%
Equation (\ref{7}) stipulates that evolution over time of per capita
consumption also depends upon the level of per capita real financial wealth $%
\overline{a}(t)$. A higher real wealth brings about a stimulus to current
consumption at expense of future consumption. This is because the absence of
perfect intergenerational altruism implies that future cohorts' consumption
is not valued by agents alive today. Government liabilities are net wealth
for living generations, for agents may not be alive to pay future taxes
required to guarantee public solvency. Intergenerational heterogeneity
prevails: older generations are wealthier than younger generations, and
therefore consume more and save less. Only in the special case in which the
birth rate $\beta $\ is equal to zero, the law of motion of per capita
consumption collapses to the traditional Euler equation characterizing the
infinitely-lived single representative agent monetary framework (e.g.,
Benhabib, Schmitt-Groh\'{e} and Uribe, 2002), whereby solely interest-rate
movements affect consumption dynamics.

\subsection{The Public Sector}

The monetary authority controls the nominal interest rate $R(t)$ and is
subject to the zero lower bound on policy rates, which prevents to set a
negative value of $R(t)$. As pointed out by Buiter (2020), the effective
lower bound that the monetary authority has to satisfy must equal the zero
nominal interest rate on currency minus the carry cost of currency, due for
instance to storage or insurance. In my model, I have abstracted from the
carry cost of currency, for analytical convenience. So the effective lower
bound must be zero. Modifying the model to account for the possibility of a
negative nominal interest rate, which is consistent with the recent behavior
of some central banks, would not alter the results of the present analysis
in any essential dimension. All that is necessary for my findings to hold is
that there exists some\ binding lower bound on policy rates. Whether such a
bound is negative, zero, or positive is inconsequential.

Having said this, let me assume that monetary policy is described by a
feedback policy rule of the form%
\begin{equation}
R(t)=\Psi (\pi (t)),  \label{taylor}
\end{equation}%
where $\Psi (\cdot )$ is a continuous, positive, increasing and strictly
convex function. The assumption of a strictly positive nominal interest rate
in the present model is made to avoid discontinuity in money demand when $%
R(t)=0$. Denoting by $\pi ^{\ast }>0$ the target level of steady state
inflation, I assume $\Psi ^{\prime }(\pi ^{\ast })>1$. That is, monetary
policy is `active' by satisfying the Taylor principle (see Taylor, 1999,
2012, 2021, Woodford, 2003, {Gal\'{\i},} 2015, and Walsh, 2017). This
constraint is meant to guarantee that whenever monetary policy makers detect
symptoms of inflationary (disinflationary) pressure, they will tighten
(ease) policy sufficiently to ensure an increase (decrease) in the real
interest rate.

Public transfer payments and interest payments on government bonds are
financed by lump-sum taxation, seignorage revenues and issuance of new
bonds. Without loss of generality for the present analysis, I set government
purchases equal to zero. Hence, the flow budget constraint of the public
sector in per capita terms is given by
\begin{equation}
\dot{\overline{b}}(t)+\dot{\overline{m}}(t)=\left( R(t)-\pi (t)-n\right)
\overline{b}(t)-\overline{\tau }(t)-\left( \pi (t)-n\right) \overline{m}(t).
\label{9}
\end{equation}%
Following Benhabib, Schmitt-Groh\'{e} and Uribe (2001, 2002) and Canzoneri,
Cumby and Diba (2010), equation (\ref{9}) can be written as%
\begin{equation}
\dot{\overline{a}}(t)=\left( R(t)-\pi (t)-n\right) \overline{a}(t)-\overline{%
s}\left( t\right) ,  \label{9a}
\end{equation}%
where $\overline{a}(t)=\overline{b}(t)+\overline{m}(t)$ are total government
liabilities and $\overline{s}\left( t\right) =\overline{\tau }(t)+R(t)%
\overline{m}(t)$ is the primary surplus inclusive of interest savings from
the issuance of money.

For my purposes, throughout the paper fiscal policies are assumed to be
Ricardian in the sense of Benhabib, Schmitt-Groh\'{e} and Uribe (2001,
2002). That is, the government must respect the terminal boundary condition
precluding Ponzi's games and requiring that the present discounted value of
total government liabilities in per capita terms converges to zero,%
\begin{equation}
\underset{t\rightarrow \infty }{\lim }\overline{a}(t)e^{-\int_{0}^{t}(R(j)-%
\pi (j)-n)dj}=0,  \label{solvency}
\end{equation}%
for all possible, equilibrium and off-equilibrium, time paths of the
remaining endogenous variables. Integrating equation (\ref{9a}) forward,
given condition (\ref{solvency}) that ensures public solvency, yields the
intertemporal budget constraint of the public sector:%
\begin{equation}
\overline{a}(0)=\int_{0}^{\infty }\overline{s}\left( t\right)
e^{-\int_{0}^{t}(R(j)-\pi (j)-n)dj}dt.  \label{IBC}
\end{equation}

\subsection{Equilibrium Inflation Dynamics}

Equilibrium in the goods' market requires that $\overline{y}\left( t\right) =%
\overline{c}(t)$. Equilibrium in the money market implies $\overline{m}(t)=%
\overline{c}(t)/\Omega (R(t))$. Total output $\overline{Y}(t)$ is assumed to
grow at the rate of population growth $n$, without loss of generality. Per
capita output is thus constant, $\overline{y}(t)=\overline{y}$.

For a generic variable at the population level $\overline{Z}(t)$, let now%
\begin{equation}
z\left( t\right) \equiv \frac{\overline{Z}\left( t\right) }{\overline{Y}%
\left( t\right) }=\frac{\overline{z}\left( t\right) }{\overline{y}}
\label{definition-output-ratios}
\end{equation}%
indicate the corresponding variable relative to the size of the economy $%
\overline{Y}\left( t\right) $. It then follows
\begin{equation}
m(t)=\frac{1}{\Omega (R(t))}.  \label{equilibrium-money}
\end{equation}%
Furthermore, from the law of motion of per capita consumption (\ref{7}), the
real interest rate that guarantees equilibrium in the goods' market is
increasing in the real value of government liabilities relative to output:%
\begin{equation}
R(t)-\pi (t)=\rho +\frac{\Lambda ^{\prime }(R(t))}{\Lambda (R(t))}\dot{R}(t)+%
\frac{\beta (\rho +\mu )}{\Lambda (R(t))}a\left( t\right) .  \label{r}
\end{equation}%
Using the monetary policy rule (\ref{taylor}) into equation (\ref{r}), one
obtains that equilibrium dynamics of inflation must obey%
\begin{equation}
\dot{\pi}(t)=\frac{\Lambda (\Psi (\pi (t)))}{\Lambda ^{\prime }(\Psi (\pi
(t)))\Psi ^{\prime }(\pi (t))}\left( \Psi (\pi (t))-\pi (t)-\rho \right) -%
\frac{\beta (\rho +\mu )}{\Lambda ^{\prime }(\Psi (\pi (t)))\Psi ^{\prime
}(\pi (t))}a\left( t\right) .  \label{11}
\end{equation}%
From (\ref{11}), the time path of inflation is\ influenced by the real
financial wealth-to-output ratio, except in the limiting case in which the
birth rate $\beta $ is equal to zero. Therefore, in order to close the model
and investigate the implied macroeconomic dynamics, one needs to specify the
policy rule describing the behavior of the fiscal authority.

\section{Strict Fiscal Discipline and Liquidity Traps}

Let me first pay attention to the dynamic consequences of a budgetary policy
regime in which the fiscal authority gradually adjusts the stock of real
government liabilities relative to the size of the economy in order to
converge towards a given target level $a^{\ast }>0$ in the long run. Thus,
consistently with Minea and Villieu (2013), Maebayashi, Hori and Futagami
(2017) and Cheron et al. (2019), the fiscal adjustment rule takes the form%
\begin{equation}
\dot{a}(t)=-\phi \left( a(t)-a^{\ast }\right) ,
\label{fiscal-adjustment-rule}
\end{equation}%
where $a^{\ast }$ is to be interpreted as a government policy parameter and $%
\phi >0$ captures the pace of the consolidation in the case in which $a(t)$
is larger than $a^{\ast }$.

Given the supply of money that in equilibrium endogenously adjusts to the
demand of money according to condition (\ref{equilibrium-money})---since in
my model the monetary authority controls the nominal interest rate $R(t)$ on
the basis of rule (\ref{taylor})---equation (\ref{fiscal-adjustment-rule})
consequently sets down the equilibrium issuance of government bonds, given by%
\begin{equation}
b\left( t\right) =a(t)-\frac{1}{\Omega (\Psi (\pi (t)))}.
\label{bonds-issuance}
\end{equation}%
Combining (\ref{9a}) expressed in per unit of output and (\ref%
{fiscal-adjustment-rule}), it also follows that to implement the fiscal
policy targeting rule the government must adjust the primary surplus
according to
\begin{equation}
s\left( t\right) =\left( \Psi (\pi (t))+\phi -\pi (t)-n\right) a(t)-\phi
a^{\ast }.  \label{ps}
\end{equation}

Setting $\dot{\pi}(t)=0$ and $\dot{a}(t)=0$ in equations (\ref{11}) and (\ref%
{fiscal-adjustment-rule}) yields the following `modified Fisher equation':%
\begin{equation}
\Psi (\pi )=\rho +\frac{\beta (\rho +\mu )}{\Lambda (\Psi (\pi ))}a^{\ast
}+\pi .  \label{modified-fisher}
\end{equation}%
Because the monetary policy reaction function $\Psi (\cdot )$ is
positive---respecting the zero lower bound on nominal interest rates---and
satisfies $\Psi ^{\prime }(\cdot ),\Psi ^{\prime \prime }(\cdot )>0$, the
steady-state relation (\ref{modified-fisher}) has two solutions, $\pi ^{\ast
}$ and $\pi ^{L}$. In addition, since I have assumed that $\pi ^{\ast }>0$
is the target inflation rate at which $\Psi ^{\prime }(\pi ^{\ast })>1$, the
alternative steady-state value $\pi ^{L}$ must obey $\pi ^{L}<\pi ^{\ast }$,
is possibly negative, and necessarily does not verify the Taylor principle,
that is, $\Psi ^{\prime }(\pi ^{L})<1$, so that monetary policy is
`passive'. Thus, the Taylor principle cannot prevail globally.

Examine next local equilibrium dynamics. Linearizing the dynamic equation (%
\ref{11}) around a steady-state point $\left( a^{\ast },\pi \right) $ and
using (\ref{fiscal-adjustment-rule}), one obtains the system%
\begin{equation}
\left(
\begin{array}{c}
\dot{a}(t) \\
\dot{\pi}(t)%
\end{array}%
\right) =J^{\left( a^{\ast },\pi \right) }\left(
\begin{array}{c}
a(t)-a^{\ast } \\
\pi (t)-\pi%
\end{array}%
\right) ,  \label{matrix}
\end{equation}%
where
\begin{equation}
J^{\left( a^{\ast },\pi \right) }=\left(
\begin{array}{cc}
-\phi & 0 \\
-\frac{\beta (\rho +\mu )}{\Lambda ^{\prime }\left( \Psi (\pi )\right) \Psi
^{\prime }(\pi )} & J_{22}^{\left( a^{\ast },\pi \right) }%
\end{array}%
\right) ,  \label{J}
\end{equation}%
with%
\begin{equation*}
J_{22}^{\left( a^{\ast },\pi \right) }=\frac{\left( \Psi ^{\prime }(\pi
)-1\right) \Lambda \left( \Psi (\pi )\right) }{\Lambda ^{\prime }\left( \Psi
(\pi )\right) \Psi ^{\prime }(\pi )}+\frac{\beta (\rho +\mu )}{\Lambda
\left( \Psi (\pi )\right) }a^{\ast }.
\end{equation*}%
The two eigenvalues of the Jacobian matrix $J^{\left( a^{\ast },\pi \right)
} $ are $-\phi <0$ and $J_{22}^{\left( a^{\ast },\pi \right) }$. Observe
that $J_{22}^{\left( a^{\ast },\pi ^{\ast }\right) }>0$ since $\Psi ^{\prime
}(\pi ^{\ast })>1$, so that one eigenvalue is positive and one eigenvalue is
negative when evaluated at the target-inflation steady state $\left( a^{\ast
},\pi ^{\ast }\right) $. Because $\pi (t)$ is a jump variable with a free
initial condition and $a\left( t\right) $ is a state variable,\footnote{%
The stock of real government liabilities relative to output should be
counted as a state variable of the system because its value cannot jump
independently of the inflation rate. To see this, denote by $A\left(
0\right) $ and $M\left( 0\right) $ the initial stocks of nominal government
liabilities and nominal money, respectively, whose values are predetermined.
Hence, the ratio $A\left( 0\right) /M\left( 0\right) =a\left( 0\right)
/m\left( 0\right) =a\left( 0\right) \Omega (\Psi (\pi (0)))$ cannot jump,
because $A\left( 0\right) /M\left( 0\right) $ is predetermined. It follows
that only $\pi (0)$ can jump freely in system (\ref{matrix}), and the
Blanchard-Kahn conditions guarantee that a steady state is locally
determined if one root of the Jacobian matrix is positive and one root is
negative.} local determinacy of equilibrium prevails in the neighborhood of
the intended steady state. That is, around $\left( a^{\ast },\pi ^{\ast
}\right) $ there exists a unique equilibrium converging asymptotically to
that steady state. Specifically, the only trajectory of $\left( a\left(
t\right) ,\pi (t)\right) $ converging asymptotically to $\left( a^{\ast
},\pi ^{\ast }\right) $ is given by the following saddle-path solution:%
\begin{equation}
\pi \left( t\right) =\pi ^{\ast }+\frac{\beta (\rho +\mu )}{\Lambda ^{\prime
}\left( \Psi (\pi ^{\ast })\right) \Psi ^{\prime }(\pi ^{\ast })\left( \phi
+J_{22}^{\left( \pi ^{\ast },a^{\ast }\right) }\right) }\left( a\left(
t\right) -a^{\ast }\right) ,  \label{transitional-adjustment1}
\end{equation}%
\begin{equation}
a\left( t\right) =a^{\ast }+\left( a\left( 0\right) -a^{\ast }\right)
e^{-\phi t}.  \label{transitional-adjustment2}
\end{equation}%
where equation (\ref{transitional-adjustment1}) describes the stable arm of
the saddle path, which is positively sloped---reflecting the fact that real
government liabilities positively affect the inflation rate via the wealth
effect on consumption. Observe, on the other hand, that $J_{22}^{\left(
a^{\ast },\pi ^{L}\right) }\gtrless 0$ if and only if $\left[ \beta (\rho
+\mu )/\Lambda \left( \Psi (\pi )\right) \right] a^{\ast }\gtrless \left(
1-\Psi ^{\prime }(\pi ^{L})\right) \Lambda \left( \Psi (\pi ^{L})\right) /%
\left[ \Lambda ^{\prime }\left( \Psi (\pi ^{L})\right) \Psi ^{\prime }(\pi
^{L})\right] $, since $\Psi ^{\prime }(\pi ^{L})<1$. Appendix C shows, in
particular, that the case $J_{22}^{\left( a^{\ast },\pi ^{L}\right) }<0$
proves to largely apply for any empirically plausible model's
parameterization. So both eigenvalues are robustly negative when evaluated
at the low-inflation steady state $\left( a^{\ast },\pi ^{L}\right) $. This
implies that local indeterminacy of equilibrium prevails in the neighborhood
of the unintended steady state. That is, around $\left( a^{\ast },\pi
^{L}\right) $ there exists a continuum of equilibrium paths of $\left(
a\left( t\right) ,\pi (t)\right) $ converging asymptotically to that steady
state.

Now, one might arguably appeal to the equilibrium determinacy result around
the intended steady state to persuade readers about the stabilizing effects
of strict fiscal constraints of the type given by the adjustment rule (\ref%
{fiscal-adjustment-rule}), in particular by advocating three properties of
the local solution. First, under the debt targeting fiscal regime equations (%
\ref{transitional-adjustment1}) and (\ref{transitional-adjustment2}) imply
that, for $a\left( 0\right) \neq a^{\ast }$, inflation uniquely converges
towards the central bank's target rate. Second, the higher the speed of
adjustment of government liabilities towards the long-run objective $a^{\ast
}$, measured by the parameter $\phi $, the faster inflation approaches to
the target. Third, equations (\ref{modified-fisher}) and (\ref%
{transitional-adjustment1})-(\ref{transitional-adjustment2}) yield the
following long-run and impact effects of a government policy change in the
target level $a^{\ast }$ on the inflation rate:
\begin{equation}
\frac{d\pi ^{\ast }}{da^{\ast }}=\frac{\beta (\rho +\mu )}{\Lambda (\Psi
(\pi ^{\ast }))\left( \Psi ^{\prime }(\pi ^{\ast })-1\right) +\beta (\rho
+\mu )\Lambda ^{\prime }(\Psi (\pi ^{\ast }))\Psi ^{\prime }(\pi ^{\ast
})/\Lambda (\Psi (\pi ^{\ast }))}>0,  \label{steady-state-effect}
\end{equation}%
\begin{equation}
\frac{d\pi \left( 0\right) ^{+}}{da^{\ast }}=\frac{d\pi ^{\ast }}{da^{\ast }}%
+\frac{\beta (\rho +\mu )}{\Lambda ^{\prime }\left( \Psi (\pi ^{\ast
})\right) \Psi ^{\prime }(\pi ^{\ast })\left( \phi +J_{22}\right) }\left(
\frac{da\left( 0\right) ^{+}}{d\pi \left( 0\right) ^{+}}\frac{d\pi \left(
0\right) ^{+}}{da^{\ast }}-1\right) ,  \label{impact1}
\end{equation}%
so that, using the fact that $A\left( 0\right) /M\left( 0\right) =a\left(
0\right) /m\left( 0\right) =a\left( 0\right) \Omega (\Psi (\pi (0)))$, where
$A\left( 0\right) $ and $M\left( 0\right) $ denote the initial predetermined
stocks of nominal government liabilities and nominal money, respectively,
which implies $da\left( 0\right) ^{+}/d\pi \left( 0\right) ^{+}=-\Omega
^{\prime }(\Psi (\pi (0)))\Psi ^{\prime }(\pi (0)/$ $\left[ \left( A\left(
0\right) /M\left( 0\right) \right) \Omega (\Psi (\pi (0)))^{2}\right] $,
\begin{eqnarray}
\frac{d\pi \left( 0\right) ^{+}}{da^{\ast }} &=&\frac{\phi \beta (\rho +\mu )%
}{\Lambda ^{\prime }\left( \Psi (\pi ^{\ast })\right) \Psi ^{\prime }(\pi
^{\ast })J_{22}\left( \phi +J_{22}\right) }  \notag \\
&&\times \left( 1+\frac{\beta (\rho +\mu )}{\Lambda ^{\prime }\left( \Psi
(\pi ^{\ast })\right) \Psi ^{\prime }(\pi ^{\ast })\left( \phi
+J_{22}\right) }\frac{\Omega ^{\prime }(\Psi (\pi (0)))\Psi ^{\prime }(\pi
(0)}{\frac{A\left( 0\right) }{M\left( 0\right) }\Omega (\Psi (\pi (0)))^{2}}%
\right) ^{-1}  \notag \\
&>&0.  \label{impact-eff}
\end{eqnarray}%
Thus, a reduction in the target level of real government liabilities
relative to the size of the economy enables the monetary authority to set a
lower inflation target, for it induces a fall in the steady-state real
interest rate via the modified Fisher equation. This reflects the fact that
the fiscal consolidation brings about a redistribution of resources from
current to future generations, since it lowers the future tax burden
required to finance the reduced interest payments on government bonds. On
impact inflation jumps downwards from $\pi (0)$ to $\pi \left( 0\right) ^{+}$
onto the new stable arm described by equation (\ref{transitional-adjustment1}%
), undershooting the new long-run equilibrium, since $d\pi \left( 0\right)
^{+}/da^{\ast }<d\pi ^{\ast }/da^{\ast }$. Notice that only in the special
case of a zero birth rate, $\beta =0$, inflation would be independent of
fiscal variables, since the model would collapse into the infinitely-lived
single representative agent paradigm, exhibiting the Ricardian debt
equivalence and no longer describing an economy with intergenerational
heterogeneity.

Summing up, should the economy be expected to permanently remain in the
neighborhood of the inflation target, the above three implications obtained
in the general case $\beta >0$ do support the consensus view that an
appropriate fiscal commitment and a correct pace at which a fiscal
consolidation plan is implemented sustain the central bank's objective of
maintaining price stability. The point, however, is that the overall story
is radically different if one does not limit to local analysis and look
instead at the global behavior of the system.

To this purpose, first observe that from equation (\ref%
{fiscal-adjustment-rule}), the $\dot{a}(t)=0$--locus is given by $a\left(
t\right) =a^{\ast }$, which in the phase plane $\left( a\left( t\right) ,\pi
(t)\right) $ is vertical. On the other hand, from equation (\ref{11}), the $%
\dot{\pi}(t)=0$--locus is implicitly given by%
\begin{equation}
\left( \Psi (\pi (t))-\pi (t)-\rho \right) \Lambda (\Psi (\pi (t)))=\beta
(\rho +\mu )a\left( t\right) .  \label{isocline-p}
\end{equation}%
Because%
\begin{equation}
\left. \frac{d\pi (t)}{da\left( t\right) }\right \vert _{\dot{\pi}(t)=0}=%
\frac{\beta (\rho +\mu )}{\left( \Psi ^{\prime }(\pi (t))-1\right) \Lambda
(\Psi (\pi (t)))+\left( \Psi (\pi (t))-\pi (t)-\rho \right) \Lambda ^{\prime
}\left( \Psi (\pi (t))\right) \Psi ^{\prime }(\pi (t))},
\label{slope-isocline-p}
\end{equation}%
which is positive at $\left( a^{\ast },\pi ^{\ast }\right) $, negative at $%
\left( a^{\ast },\pi ^{L}\right) $ (see Appendix C), and infinity when $%
\left( 1-\Psi ^{\prime }(\pi (t))\right) \Lambda (\Psi (\pi (t)))=\left(
\Psi (\pi (t))-\pi (t)-\rho \right) \Lambda ^{\prime }\left( \Psi (\pi
(t))\right) \Psi ^{\prime }(\pi (t))$, in the phase plane $\left( a\left(
t\right) ,\pi (t)\right) $ the $\dot{\pi}(t)=0$--locus is horizontally
U-shaped. The two loci intersect at the steady states $\left( a^{\ast },\pi
^{\ast }\right) $ and $\left( a^{\ast },\pi ^{L}\right) $. Then, from (\ref%
{fiscal-adjustment-rule}) and (\ref{11}), I have $\dot{a}(t)>\left( <\right)
0$ if $a<\left( >\right) a^{\ast }$ and $\dot{\pi}(t)>\left( <\right) 0$ if $%
\left( \Psi (\pi (t))-\pi (t)-\rho \right) \Lambda (\Psi (\pi (t)))>\left(
<\right) \beta (\rho +\mu )a\left( t\right) $.

The resulting global dynamics are characterized in Figure 1.
\begin{figure}[t]
\begin{center}
\includegraphics[scale=0.45]{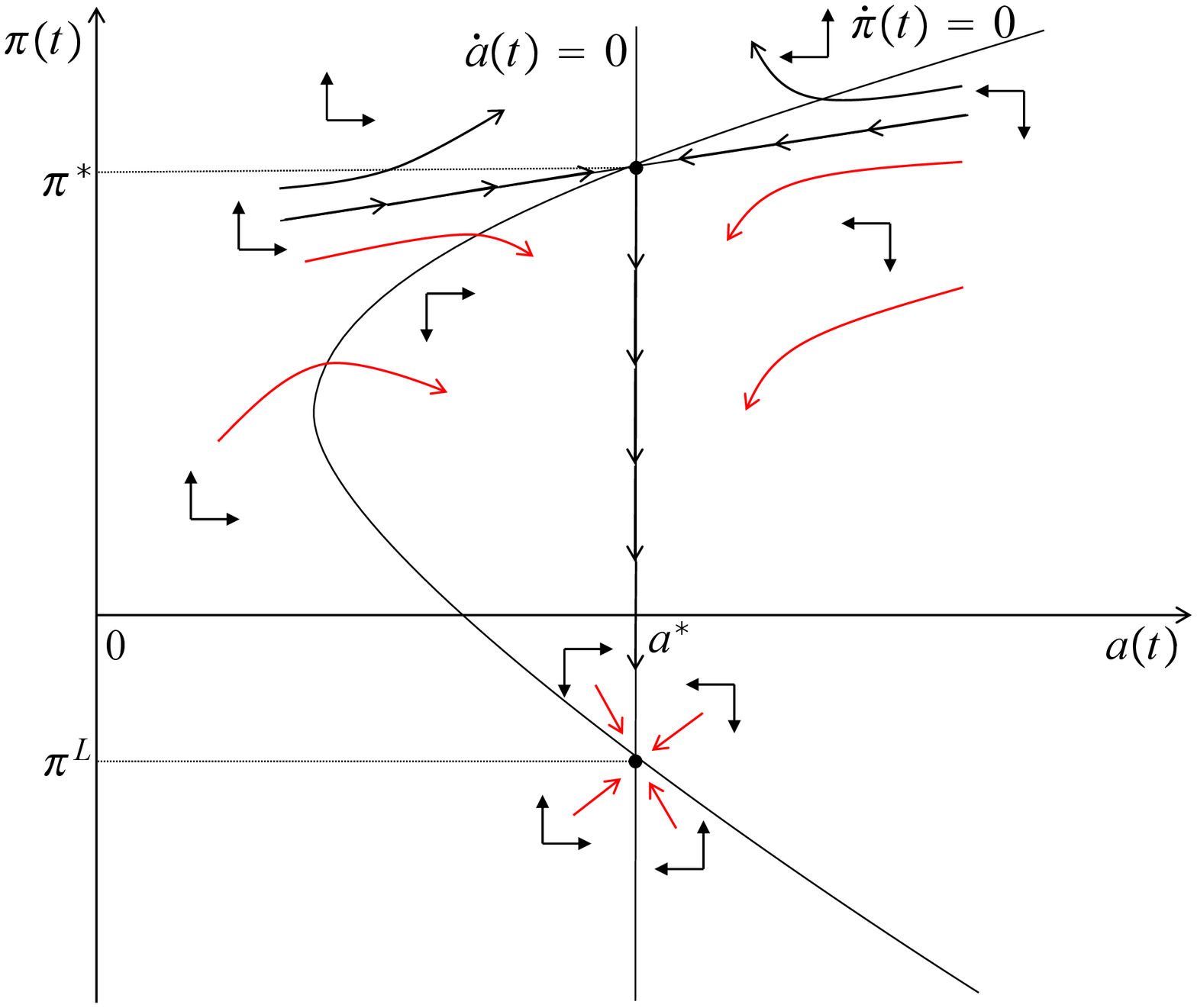}
\end{center}
\caption{\textit{Dynamic behavior of }$\left( a\left( t\right) ,\protect \pi %
(t)\right) $\textit{\ under strict fiscal discipline}}
\end{figure}
The stable arm of the saddle point passing through the steady state $\left(
a^{\ast },\pi ^{\ast }\right) $, has a slope given by $\beta (\rho +\mu )/%
\left[ \Lambda ^{\prime }\left( \Psi (\pi ^{\ast })\right) \Psi ^{\prime
}(\pi ^{\ast })\left( \phi +J_{22}^{\left( \pi ^{\ast },a^{\ast }\right)
}\right) \right] >0$ that is lower than the slope of the $\dot{\pi}(t)=0$%
--locus evaluated at $\left( a^{\ast },\pi ^{\ast }\right) $, given by $%
\beta (\rho +\mu )/\left[ \Lambda ^{\prime }\left( \Psi (\pi ^{\ast
})\right) \Psi ^{\prime }(\pi ^{\ast })J_{22}^{\left( \pi ^{\ast },a^{\ast
}\right) }\right] >0$. It clearly emerges that, in the neighborhood of the
intended steady state $\left( a^{\ast },\pi ^{\ast }\right) $, there exists
an infinite number of initial values for the inflation rate featured by $\pi
\left( 0\right) <\pi ^{\ast }+$ $\left \{ \beta (\rho +\mu )/\left[ \Lambda
^{\prime }\left( \Psi (\pi ^{\ast })\right) \Psi ^{\prime }(\pi ^{\ast
})\left( \phi +J_{11}^{\left( \pi ^{\ast },a^{\ast }\right) }\right) \right]
\right \} \left( a\left( 0\right) -a^{\ast }\right) $ such that $\left(
a\left( t\right) ,\pi (t)\right) $ will spiral into deflationary
trajectories converging to the trapping steady state $\left( a^{\ast },\pi
^{L}\right) $.\footnote{%
Notice that global inflation dynamics originating around $\left( a^{\ast
},\pi ^{\ast }\right) $ follow a (non-)monotonic behavior if they start to
the right (left) of the $\dot{\pi}(t)=0$--locus.} The saddle manifold
associated with $\left( a^{\ast },\pi ^{\ast }\right) $ is precisely the
boundary of the basin of attraction of $\left( a^{\ast },\pi ^{L}\right) $.

In other words, once a global perspective is accounted for, inflation no
longer needs to jump on the saddle path leading to the target rate to
guarantee dynamic stability. All initial values of inflation bounded above
by the saddle path are equilibrium values for they make $\left( a\left(
t\right) ,\pi (t)\right) $ approach to the alternative steady state $\left(
a^{\ast },\pi ^{L}\right) $. This implies that the economic system is
affected by global indeterminacy.

In the present framework with dyachronous heterogeneity, deflationary paths
generate, \textit{per se}, a redistribution of real financial wealth from
future to current generations, since the associated rise in the stock of
government liabilities in real terms causes the fiscal burden borne by
future cohorts' to increase in order to avoid public insolvency. This
intergenerational reallocation of resources beneficial to current
generations would exert, \textit{ceteris paribus}, a stimulus to aggregate
consumption and as a consequence to prices, thereby potentially offsetting
the initial deflationary perturbations, along lines analogous to the
traditional wealth-effect channel emphasized in the static models of Pigou
(1943, 1947) and Patinkin (1965). Nevertheless, if the fiscal policy regime
forces the government to adjust gradually the real amount of government
liabilities relative to the size of the economy at a certain
level---similarly, say, to the permanent commitment rules in the Maastricht
Treaty and the Fiscal Compact reforming the Stability and the Growth Pact in
the European Union\footnote{%
The Fiscal Compact---formally `Treaty on Stability, Coordination and
Governance in the Economic and Monetary Union'---is in force since 2013,
although temporarily suspended from 2020 to 2023 because of the pandemic
crisis, and features a well-defined debt reduction benchmark rule.
Specifically, the rule establishes that Member States whose debt-to-GDP
ratio exceeds the 60\% threshold are required to reduce their ratios to the
reference value at an average rate of one-twentieth per year.}---following
out-of-fundamentals deflationary slumps primary surpluses are expected to
expand in order to choke off a potential escalation of real debt. The
resulting fiscal austerity actions bring about a negative wealth effect on
current cohorts' consumption and consequently strengthen the deflationary
pressures over time, hence validating the initial arbitrary revision in
agents' expectations and making the economy descend in a self-fulfilling
manner into a steady state that displays the characteristics of a liquidity
trap: monetary policy turns to be powerless in positively affecting
aggregate demand and thus prices, losing control over inflation and
incurring sunspot fluctuations to the extent that the interest-rate policy
stance is passive in the vicinity of the effective lower bound---so that the
Taylor principle is no longer satisfied.

In synthesis, the stabilizing effects of fiscal consolidation strategies
that result from a local-dynamics perspective break down, and are even
reversed, from a global-dynamics perspective. Unlike conventional wisdom,
strict fiscal discipline does not support price stability for, on the
contrary, is likely to give rise to a self-fulfilling austerity-decelerating
inflation nexus that amplifies the possibility of expectations-driven
liquidity traps.

\section{Escaping Liquidity Traps through Activist Fiscal Policy Regimes}

Consider now the dynamic implications of a policy regime in which the fiscal
authorities are assumed to react actively---in a complementary way with
respect to monetary policy---in the event that off-target decelerating
inflation paths leading to a liquidity trap materialize, provided that
intertemporal budget constraint of the government is always satisfied.

Applying (\ref{definition-output-ratios}) to (\ref{9a}) yields the flow
budget constraint of the public sector in terms of ratios to output:%
\begin{equation}
\dot{a}(t)=\left( R(t)-\pi (t)-n\right) a(t)-s\left( t\right) .
\label{budget-output-ratios}
\end{equation}%
The solvency condition thus requires that%
\begin{equation}
\underset{t\rightarrow \infty }{\lim }a(t)e^{-\int_{0}^{t}(R(j)-\pi
(j)-n)dj}=0,  \label{solvency-output-ratio}
\end{equation}%
or, equivalently,%
\begin{equation}
a(0)=\int_{0}^{\infty }s\left( t\right) e^{-\int_{0}^{t}(R(j)-\pi
(j)-n)dj}dt.  \label{IBC-output-ratios}
\end{equation}%
Suppose, in particular, that the feedback budgetary policy rule takes the
form%
\begin{equation}
s\left( t\right) =\Theta \left( \Psi (\pi (t))-\pi (t)\right) a\left(
t\right) +\Gamma \left( \Psi (\pi (t))-\pi (t)\right) ,
\label{activist-fiscal-rule}
\end{equation}%
where $\Theta (\cdot )$ and $\Gamma (\cdot )$ are continuous and increasing
functions. The idea behind rule (\ref{activist-fiscal-rule}) is that,
whenever fiscal policy makers observe symptoms of decelerating
inflation---with the central bank that decreases the real interest rate
according to the Taylor principle in the attempt to reverse dynamics, they
will ease budgetary policy by ensuring a decline in primary surpluses or a
rise in primary deficits through tax cuts and/or transfer increases. The
fiscal expansion complementary to the Taylor-type feedback rule is assumed
to occur both directly via $\Gamma (\cdot )$ and indirectly via the
reduction of the sensitivity of the primary surplus with respect to total
government liabilities, measured by $\Theta (\cdot )$. Functions $\Theta
(\cdot )$ and $\Gamma (\cdot )$ are such that the solvency condition (\ref%
{solvency-output-ratio}) is globally satisfied. Specifically, using the
resulting law of motion of $a(t)$,%
\begin{equation}
\dot{a}(t)=\left( \Psi (\pi (t))-\pi (t)-n-\Theta \left( \Psi (\pi (t))-\pi
(t)\right) \right) a(t)-\Gamma \left( \Psi (\pi (t))-\pi (t)\right) ,
\label{law-of-motion-a}
\end{equation}%
this occurs when, should government liabilities embark on potentially
explosive paths, beyond a certain arbitrarily chosen threshold level for $%
a(t)$ function $\Theta (\cdot )$ becomes permanently positive (see, e.g.,
Bohn, 1991, 1998), so that%
\begin{equation}
\underset{t\rightarrow \infty }{\lim }\int_{0}^{t}\Theta \left( \Psi (\pi
(v))-\pi (v)\right) dv=+\infty ,  \label{solvency1}
\end{equation}%
and, in addition,%
\begin{equation}
\underset{t\rightarrow \infty }{\lim }-e^{-\int_{0}^{t}\Theta \left( \Psi
(\pi (v))-\pi (v)\right) dv}\int_{0}^{t}\Gamma \left( \Psi (\pi (v))-\pi
(v)\right) e^{-\int_{0}^{v}(R(j)-\pi (j)-n-\Theta \left( \Psi (\pi (j))-\pi
(j)\right) )dj}dv=0.  \label{solvency2}
\end{equation}%
Under (\ref{solvency1}), condition (\ref{solvency2}) is surely verified when
the present discounted value of future primary deficits associated to
function $\Gamma (\cdot )$ in (\ref{activist-fiscal-rule}) is lower than $%
+\infty $---an arguably realistic assumption.

Setting $\dot{a}(t)=0$ and $\dot{\pi}(t)=0$ in the dynamic equations (\ref%
{law-of-motion-a}) and (\ref{11}) yields%
\begin{equation}
a=\frac{\Gamma \left( \Psi (\pi )-\pi \right) }{\left( \Psi (\pi )-\pi
-n-\Theta \left( \Psi (\pi )-\pi \right) \right) },  \label{a-dot=0}
\end{equation}%
\begin{equation}
\Psi (\pi )=\rho +\frac{\beta (\rho +\mu )}{\Lambda (\Psi (\pi ))}a+\pi .
\label{p-dot=0}
\end{equation}%
Close inspection of (\ref{a-dot=0}) and (\ref{p-dot=0}) reveals that there
exist at least two steady states, $\left( a^{\ast },\pi ^{\ast }\right) $
and $\left( a^{L},\pi ^{L}\right) $, obeying $\pi ^{\ast }>\pi ^{L}$, $\Psi
^{\prime }(\pi ^{\ast })>1$, $\Psi ^{\prime }(\pi ^{L})<1$, and $a^{\ast
}\gtreqless a^{L}$. This reflects the fact that, for a given $a$, the
modified Fisher equation (\ref{p-dot=0}) continues to have two solutions for
$\pi $, one in which monetary policy is active and one in which monetary
policy is passive, analogously to the previous section.

Linearizing equations (\ref{law-of-motion-a}) and (\ref{11}) in the
neighborhood of any steady-state point $\left( a,\pi \right) $, one obtains
the system%
\begin{equation}
\left(
\begin{array}{c}
\dot{a}(t) \\
\dot{\pi}(t)%
\end{array}%
\right) =K^{\left( a,\pi \right) }\left(
\begin{array}{c}
a(t)-a \\
\pi (t)-\pi
\end{array}%
\right) ,  \label{matrix1}
\end{equation}%
where
\begin{equation}
K^{\left( a,\pi \right) }=\left(
\begin{array}{cc}
\Psi (\pi )-\pi -n-\Theta \left( \Psi (\pi )-\pi \right)  & \left( \Psi
^{\prime }(\pi )-1\right) \left(
\begin{array}{c}
a-\Theta ^{\prime }\left( \Psi (\pi )-\pi \right) a \\
-\Gamma ^{\prime }\left( \Psi (\pi )-\pi \right)
\end{array}%
\right)  \\
-\frac{\beta (\rho +\mu )}{\Lambda ^{\prime }\left( \Psi (\pi )\right) \Psi
^{\prime }(\pi )} & K_{22}^{\left( a,\pi \right) }%
\end{array}%
\right) ,  \label{K}
\end{equation}%
with%
\begin{equation*}
K_{22}^{\left( a,\pi \right) }=\frac{\left( \Psi ^{\prime }(\pi )-1\right)
\Lambda \left( \Psi (\pi )\right) }{\Lambda ^{\prime }\left( \Psi (\pi
)\right) \Psi ^{\prime }(\pi )}+\frac{\beta (\rho +\mu )}{\Lambda \left(
\Psi (\pi )\right) }a.
\end{equation*}%
The determinant and the trace of the Jacobian matrix $K^{\left( a,\pi
\right) }$ are%
\begin{eqnarray}
\det \text{ }K^{\left( a,\pi \right) } &=&\left( \Psi (\pi )-\pi -n-\Theta
\left( \Psi (\pi )-\pi \right) \right) K_{22}^{\left( a,\pi \right) }  \notag
\\
&&+\frac{\beta (\rho +\mu )}{\Lambda ^{\prime }\left( \Psi (\pi )\right)
\Psi ^{\prime }(\pi )}\left( \Psi ^{\prime }(\pi )-1\right) \left(
\begin{array}{c}
a-\Theta ^{\prime }\left( \Psi (\pi )-\pi \right) a \\
-\Gamma ^{\prime }\left( \Psi (\pi )-\pi \right)
\end{array}%
\right) ,  \label{det}
\end{eqnarray}%
\begin{equation}
\text{tr }K^{\left( a,\pi \right) }=\Psi (\pi )-\pi -n-\Theta \left( \Psi
(\pi )-\pi \right) +K_{22}^{\left( a,\pi \right) }.  \label{tr}
\end{equation}%
Observe that $\det $ $K^{\left( a^{\ast },\pi ^{\ast }\right) }<0$ if%
\begin{eqnarray}
&&\Gamma ^{\prime }\left( \Psi (\pi ^{\ast })-\pi ^{\ast }\right) +\Theta
^{\prime }\left( \Psi (\pi ^{\ast })-\pi ^{\ast }\right) a^{\ast }  \notag \\
&>&a^{\ast }+\frac{\left( \Psi (\pi ^{\ast })-\pi ^{\ast }-n-\Theta \left(
\Psi (\pi ^{\ast })-\pi ^{\ast }\right) \right) \Lambda ^{\prime }\left(
\Psi (\pi ^{\ast })\right) \Psi ^{\prime }(\pi ^{\ast })K_{22}^{\left(
a^{\ast },\pi ^{\ast }\right) }}{\beta (\rho +\mu )\left( \Psi ^{\prime
}(\pi ^{\ast })-1\right) }.  \label{activist-1}
\end{eqnarray}%
That is, local determinacy of equilibrium applies in the neighborhood of the
intended steady state $\left( a^{\ast },\pi ^{\ast }\right) $, where $\Psi
^{\prime }(\pi ^{\ast })>1$, provided that fiscal policy is sufficiently
reactive to potential disinflationary pressures. In particular, the locally
unique trajectory of $\left( a\left( t\right) ,\pi (t)\right) $ converging
asymptotically to $\left( a^{\ast },\pi ^{\ast }\right) $ is given by the
saddle-path solution expressed by%
\begin{equation}
\pi \left( t\right) =\pi ^{\ast }-\frac{\beta (\rho +\mu )}{\Lambda ^{\prime
}\left( \Psi (\pi ^{\ast })\right) \Psi ^{\prime }(\pi ^{\ast })\left(
\varepsilon _{1}-K_{22}^{\left( \pi ^{\ast },a^{\ast }\right) }\right) }%
\left( a\left( t\right) -a^{\ast }\right) ,
\label{transitional-adjustment-3}
\end{equation}%
\begin{equation}
a\left( t\right) =a^{\ast }+\left( a\left( 0\right) -a^{\ast }\right)
e^{\varepsilon _{1}t}.  \label{transitional-adjustment-4}
\end{equation}%
where $\varepsilon _{1}$ is the negative eigenvalue associated to $K^{\left(
a^{\ast },\pi ^{\ast }\right) }$ and equation (\ref%
{transitional-adjustment-3}) is the stable arm of the saddle path, which is
positively sloped since $K_{22}^{\left( \pi ^{\ast },a^{\ast }\right) }>0$%
---reflecting again the positive comovement between agents' wealth and
inflation. On the other hand, the unintended steady state $\left( a^{L},\pi
^{L}\right) $ with $\Psi ^{\prime }(\pi ^{L})<1$, exhibiting deflation or
disinflation, may no longer be a sink---differently from the case of a
strict fiscal discipline examined in Section 4. Indeed, observe that $\det $
$K^{\left( a^{L},\pi ^{L}\right) }>0$ if%
\begin{eqnarray}
&&\Gamma ^{\prime }\left( \Psi (\pi ^{L})-\pi ^{L}\right) +\Theta ^{\prime
}\left( \Psi (\pi ^{L})-\pi ^{L}\right) a^{L}  \notag \\
&>&a^{L}+\frac{\left( \Psi (\pi ^{L})-\pi ^{L}-n-\Theta \left( \Psi (\pi
^{L})-\pi ^{L}\right) \right) \Lambda ^{\prime }\left( \Psi (\pi
^{L})\right) \Psi ^{\prime }(\pi ^{L})\left( -K_{22}^{\left( a^{L},\pi
^{L}\right) }\right) }{\beta (\rho +\mu )\left( 1-\Psi ^{\prime }(\pi
^{L})\right) },\text{ \  \  \ }  \label{activist2}
\end{eqnarray}%
where $K_{22}^{\left( a^{L},\pi ^{L}\right) }<0$ on the basis of any
empirically plausible model's calibration (see Appendix C), and tr $%
K^{\left( a^{L},\pi ^{L}\right) }>0$ if%
\begin{equation}
\Theta \left( \Psi (\pi ^{L})-\pi ^{L}\right) <\rho -n+\frac{\beta (\rho
+\mu )}{\Lambda (\Psi (\pi ^{L}))}a+K_{22}^{\left( a^{L},\pi ^{L}\right) }.
\label{activist3}
\end{equation}%
That is, the steady state $\left( a^{L},\pi ^{L}\right) $ becomes unstable
if fiscal policy is sufficiently activist and the sensitivity of the primary
surpluses with respect to government liabilities is relatively low. For what
will follow, in particular, it is worth pointing out that $\left( \text{tr }%
K^{\left( a^{L},\pi ^{L}\right) }\right) ^{2}-4\det $ $K^{\left( a^{\ast
},\pi ^{\ast }\right) }>\left( <\right) 0$ if%
\begin{eqnarray}
&&\Gamma ^{\prime }\left( \Psi (\pi ^{L})-\pi ^{L}\right) +\Theta ^{\prime
}\left( \Psi (\pi ^{L})-\pi ^{L}\right) a^{L}  \notag \\
&<&\left( >\right) a^{L}+\left[ \frac{\left( \Psi (\pi ^{L})-\pi
^{L}-n-\Theta \left( \Psi (\pi ^{L})-\pi ^{L}\right) \right) \Lambda
^{\prime }\left( \Psi (\pi ^{L})\right) \Psi ^{\prime }(\pi ^{L})\left(
-K_{22}^{\left( a^{L},\pi ^{L}\right) }\right) }{\beta (\rho +\mu )\left(
1-\Psi ^{\prime }(\pi ^{L})\right) }\right]   \notag \\
&&+\frac{\Lambda ^{\prime }\left( \Psi (\pi ^{L})\right) \Psi ^{\prime }(\pi
^{L})\left( \Psi (\pi ^{L})-\pi ^{L}-n-\Theta \left( \Psi (\pi ^{L})-\pi
^{L}\right) +K_{22}^{\left( a^{L},\pi ^{L}\right) }\right) ^{2}}{4\beta
(\rho +\mu )\left( 1-\Psi ^{\prime }(\pi ^{L})\right) }.
\label{discriminant}
\end{eqnarray}%
As a result, provided that fiscal policy is not overly aggressive, the roots
associated to the Jacobian matrix $K^{\left( a^{L},\pi ^{L}\right) }$ are
real, implying that $\left( a^{L},\pi ^{L}\right) $ is an unstable node. In
this case, there exists a continuum of curved paths that are tangent to the
negatively-slopped unstable branch given by%
\begin{equation}
\pi \left( t\right) =\pi ^{L}-\frac{\beta (\rho +\mu )}{\Lambda ^{\prime
}\left( \Psi (\pi ^{L})\right) \Psi ^{\prime }(\pi ^{L})\left( \eta
_{1}-K_{22}^{\left( \pi ^{L},a^{L}\right) }\right) }\left( a\left( t\right)
-a^{L}\right) ,  \label{unstable-eigenspace}
\end{equation}%
where $\eta _{1}>0$ is the non-dominant eigenvalue of $K^{\left( a^{L},\pi
^{L}\right) }$. By contrast, should fiscal policy be excessively aggressive,
roots are complex, implying that $\left( a^{L},\pi ^{L}\right) $ is an
unstable spiral point.

I am now ready to elucidate the consequences for global dynamics under
activist fiscal policy regimes satisfying conditions (\ref{activist-1}), (%
\ref{activist2}) and (\ref{activist3}). From equation (\ref{law-of-motion-a}%
), the $\dot{a}(t)=0$--locus is given by%
\begin{equation}
\frac{\Gamma \left( \Psi (\pi (t))-\pi (t)\right) }{\Psi (\pi (t))-\pi
(t)-n-\Theta \left( \Psi (\pi (t))-\pi (t)\right) }=a(t),  \label{isocline-a}
\end{equation}%
from which%
\begin{equation}
\left. \frac{d\pi (t)}{da\left( t\right) }\right \vert _{\dot{a}(t)=0}=\frac{%
\Gamma \left( \Psi (\pi (t))-\pi (t)\right) }{\left( \Psi ^{\prime }(\pi
(t))-1\right) \left( \Gamma ^{\prime }\left( \Psi (\pi (t))-\pi (t)\right)
+\Theta ^{\prime }\left( \Psi (\pi (t))-\pi (t)\right) a(t)-a(t)\right) a(t)}%
.  \label{slope-locus-a}
\end{equation}%
In what follows, I shall assume the restrictions $\left. \Gamma \left( \Psi
(\pi (t))-\pi (t)\right) \right \vert _{\dot{a}(t)=0}>0$ and $\left. \Gamma
^{\prime }\left( \Psi (\pi (t))-\pi (t)\right) +\Theta ^{\prime }\left( \Psi
(\pi (t))-\pi (t)\right) a(t)-a(t)\right \vert _{\dot{a}(t)=0}>0$, according
to which fiscal policy avoids the existence of more than two steady states.
Hence, the slope of the $\dot{a}(t)=0$--locus is positive at $\left( \pi
^{\ast },a^{\ast }\right) $, negative at $\left( a^{L},\pi ^{L},\right) $,
and infinity when $\Psi ^{\prime }(\pi (t))=1$. Therefore, in the phase
plane $\left( a\left( t\right) ,\pi (t)\right) $ the $\dot{a}(t)=0$--locus
is horizontally U-shaped. As in the previous section, the $\dot{\pi}(t)=0$%
--locus implicitily given by (\ref{isocline-p}) is also horizontally
U-shaped. The two loci intersect at the steady states $\left( a^{\ast },\pi
^{\ast }\right) $ and $\left( a^{L},\pi ^{L}\right) $. From the properties
of the Jacobian matrix, it emerges that the slope of the $\dot{a}(t)=0$%
--locus is lower than the slope of the $\dot{\pi}(t)=0$--locus if evaluated
at $\left( a^{\ast },\pi ^{\ast }\right) $ and the opposite occurs if the
slopes are evaluated at $\left( a^{L},\pi ^{L}\right) $. From (\ref%
{law-of-motion-a}) and (\ref{11}), I have that $\dot{a}(t)>\left( <\right) 0$
if $\left( \Psi (\pi (t))-\pi (t)-n-\Theta \left( \Psi (\pi (t))-\pi
(t)\right) \right) a(t)>\left( <\right) \Gamma \left( \Psi (\pi (t))-\pi
(t)\right) $ and $\dot{\pi}(t)>\left( <\right) 0$ if $\left( \Psi (\pi
(t))-\pi (t)-\rho \right) \Lambda (\Psi (\pi (t)))>\left( <\right) \beta
(\rho +\mu )a\left( t\right) $.

In the case of real roots associated to $K^{\left( a^{L},\pi ^{L}\right) }$,
which require that fiscal policy is not overly aggressive at the
liquidity-trap steady state, the resulting global dynamics are characterized
in Figure 2.
\begin{figure}[t]
\begin{center}
\includegraphics[scale=0.45]{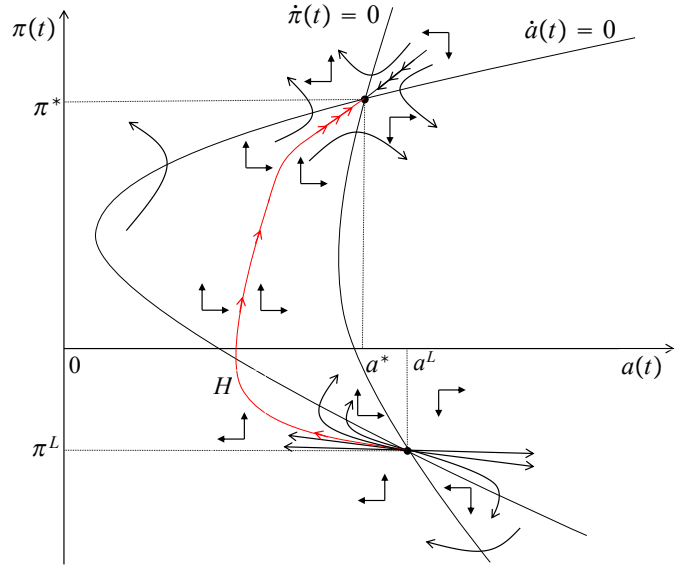}
\end{center}
\caption{\textit{Dynamic behavior of }$\left( a\left( t\right) ,\protect \pi %
(t)\right) $\textit{\ under moderately activist fiscal policies}}
\end{figure}
\begin{figure}[t]
\begin{center}
\includegraphics[scale=0.45]{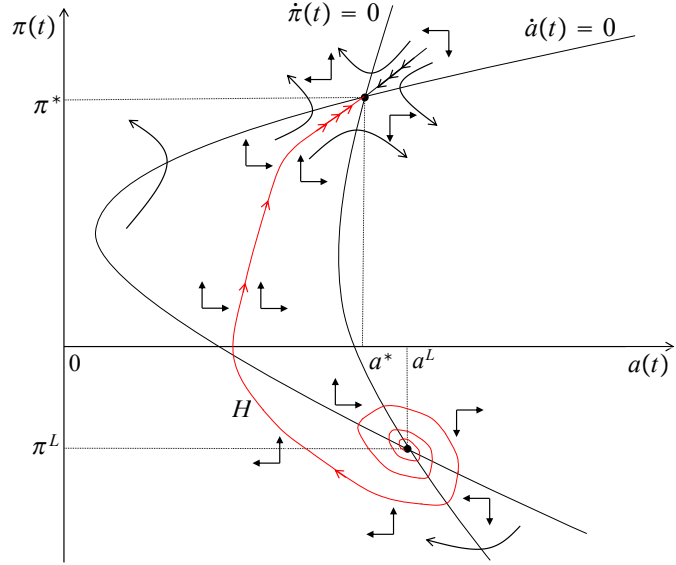}
\end{center}
\caption{\textit{Dynamic behavior of }$\left( a\left( t\right) ,\protect \pi %
(t)\right) $\textit{\ under overly activist fiscal policies}}
\end{figure}
The stable arm of the saddle point passing through $\left( a^{\ast },\pi
^{\ast }\right) $ has a positive slope given by $-\beta (\rho +\mu )/\left[
\Lambda ^{\prime }\left( \Psi (\pi ^{\ast })\right) \Psi ^{\prime }(\pi
^{\ast })\left( \varepsilon _{1}-K_{22}^{\left( \pi ^{\ast },a^{\ast
}\right) }\right) \right] >0$, which is higher than the slope of the $\dot{a}%
(t)=0$--locus evaluated at $\left( a^{\ast },\pi ^{\ast }\right) $, given by
(\ref{slope-locus-a}) at $\left( a^{\ast },\pi ^{\ast }\right) $, and lower
than the slope of the $\dot{\pi}(t)=0$--locus evaluated at $\left( a^{\ast
},\pi ^{\ast }\right) $, given by $\beta (\rho +\mu )/\left[ \Lambda
^{\prime }\left( \Psi (\pi ^{\ast })\right) \Psi ^{\prime }(\pi ^{\ast
})K_{22}^{\left( \pi ^{\ast },a^{\ast }\right) }\right] >0$. Since the
steady state $\left( a^{L},\pi ^{L}\right) $ is an unstable node, there
exists one trajectory---the heteroclinic orbit $H$---originating in the
neighborhood of the steady state $\left( a^{L},\pi ^{L}\right) $, negatively
slopped around $\left( a^{L},\pi ^{L}\right) $ because tangent to the
negatively slopped eigenspace related to the nondominant eigenvalue $\eta
_{1}$ expressed by equation (\ref{unstable-eigenspace}), and converging
asymptotically to the steady state $\left( a^{\ast },\pi ^{\ast }\right) $%
---locally along the associated saddle path whose stable arm is given by
equation (\ref{transitional-adjustment-3}).\footnote{%
As the heteroclinic orbit $H$ is tangent to the eigenspace associated to $%
\eta _{1}$ in the neighborhood $\left( a^{L},\pi ^{L}\right) $, it lies
between the $\dot{a}(t)=0$--locus and that eigenspace and never crosses this
line. As a result, the $H$--trajectory is enclosed by a `trapping region'
whose sides are given by the $\dot{\pi}(t)=0$--locus between the two
steady-state equilibria, a line passing through the steady state $\left(
a^{L},\pi ^{L}\right) $ whose slope is given by the eigenvector associated
to the non-dominant eigenvalue $\eta _{1}$, and a line passing through the
steady state $\left( a^{\ast },\pi ^{\ast }\right) $ whose slope is given by
the slope of the associated saddle path, between $\left( a^{\ast },\pi
^{\ast }\right) $ and the previous line. All the trajectories starting
inside the trapping area escape from it, with the exception of those
starting at any point along the heteroclinic orbit. In other words, only the
orbit does not hit the boundaries of the trapping area and follows a
non-monotonous path---in terms of dynamic behavior of $a(t)$---that changes
direction when $\dot{\pi}(t)=0$.} The existence of such a saddle connection
guarantees global determinacy of equilibrium, according to which even if the
economy lies in the neighborhood of the liquidity-trap steady state at which
monetary policy is passive, now inflation and real government liabilities
will \textit{uniquely} escape from $\left( a^{L},\pi ^{L}\right) $ and
converge towards the target steady state at which monetary policy is active.
The reason why liquidity traps are eradicated with \textit{no} commitment to
trigger unsustainable budget deficits is that a sequence of tax cuts and
transfer increases financed by bond issues and future primary surpluses
expands both human and financial wealth for currently alive households, thus
spuring aggregate consumption, since government debt is net wealth for
living generations. This course of sustainable fiscal policy measures, which
partly shift the sequence of future net taxes to future generations thereby
leading current generations to dissave, is capable of reflating the economy
due to the induced excess demand in the goods' market and restoring in
general equilibrium dynamic convergence of inflation towards the intended
steady state.

On the other hand, in the case of complex roots associated to $K^{\left(
a^{L},\pi ^{L}\right) }$, which occur when fiscal policy is overly
aggressive at the unintended steady state, the implied global dynamics
become those displayed in Figure 3. Even though fiscal expansions are able
to escape liquidity traps along a heteroclinic orbit leading to $\left(
a^{\ast },\pi ^{\ast }\right) $ as in the previous case, now global
indeterminacy prevails because the steady state $\left( a^{L},\pi
^{L}\right) $ is an unstable spiral point. Hence, if the economy lies in the
vicinity of the liquidity-trap steady state, there exists a large class of
initial values for the inflation rate compatible with a globally stable
perfect foresight equilibrium. Furthermore, the inflation rate may fluctuate
for relatively long periods of time in a region whereby monetary policy is
necessarily passive because of the effective lower bound on the nominal
interest rate, away from the intended steady state. Taken together, these
results give analytical support to the view that excessively lax fiscal
stimuli, although expected to be repaid by large future net taxes needed to
redeem the government debt eventually, are likely to be a quite severe
source of macroeconomic instability.

In synthesis, should the latter case of macroeconomic instability induced by
too aggressive fiscal boosts be ruled out by fiscal policy makers, the
theoretical findings demonstrated in my analysis clearly lead to the
conclusion that moderately activist demand side oriented budgetary
interventions, responding to deflationary pressures under the respect of the
government's intertemporal budget constraint, do constitute an essential
tool of `last resort' in order to avert liquidity traps and, at the same
time, preserve the stabilizing properties of the rules-based approach to
monetary policy.

\section{Conclusions}

As the nominal interest rates set by central banks in most major economies
have turned close to zero and even negative as a consequence of the secular
stagnation and in order to counter the ongoing pandemic crisis---within a
prolonged period of below-target inflation---the issue of how to escape
undesirable deflationary paths associated to a situation of liquidity trap
without creating space for other forms of macroeconomic instability, such as
fiscal unsustainability or a burst of inflation, is a pressing concern for
policy makers. The present paper explores the scope for aggregate stability
in a monetary model with overlapping generations of finitely-lived agents
that displays multiple steady-state equilibria due to the existence of a
binding lower bound on nominal interest rates. First, I depart from the
standard literature on confidence-driven liquidity traps based upon the
infinitely-lived single representative agent theoretical paradigm, because
in the present framework wealth effects on aggregate demand dynamics
emerging from intergenerational heterogeneity critically affect both the
monetary and fiscal policy transmission mechanism. Hence, the model I
present is a natural extension of the seminal work by Benhabib, Schmitt-Groh%
\'{e} and Uribe (2002). Second, I depart from the standard literature on
fiscal policy multipliers based upon local approximations of dynamic
stochastic general equilibrium models, because I take into account
non-linearities and potential connections between multiple steady-state
equilibria resulting from a global-dynamics perspective. Hence, my framework
relaxing the Ricardian equivalence due to disconnectedness across
generations is a natural alternative to the prominent work by Mertens and
Ravn (2014), grounded on a nonlinear New Keynesian model with distortionary
taxation. Third, I concentrate on the stabilizing role of intertemporal
reallocations of taxes and transfers in an otherwise standard monetary model
with flexible prices. Hence, the model I employ is a useful complement the
theoretical analysis developed by Schmidt (2016), based on the implications
of Ricardian government spending rules in the context a typical New
Keynesian model.

My analysis leads to three conclusions. First, a rigid degree of fiscal
discipline of the type prescribed by the Maastricht-Treaty and the
Fiscal-Compact frameworks in the European Union, in which the government is
required to gradually adjust the stock of real government liabilities
relative to the size of the economy in order to guarantee convergence
towards a target level in the long run, is not a precondition to price
stability, as commonly believed. On the contrary, it is a potentially severe
source of macroeconomic imbalances, because it typically makes the economy
exposed to the emergence of unintended liquidity traps, which preclude the
central bank from uniquely pinning down inflation at the target rate. The
central reason is that, under this fiscal policy regime, following
deflationary dynamics driven by arbitrary revisions in private agents'
expectations the government is expected to rise primary surpluses in order
to rule out escalation of debt in real terms. Thus austerity actions produce
a negative wealth effect on current cohorts' consumption, amplifying the
initial deflationary perturbations over time. Global indeterminacy emerges,
since any inflation trajectory bounded above by the saddle path passing
through the target steady state can be validated as an equilibrium outcome.

Second, implementing an intertemporally balanced fiscal stimulus in response
to out-of-equilibrium deflationary paths---with no need to make the
liquidity-trap steady state fiscally unsustainable as argued within the
classical Ramsey-type single representative agent paradigm---exerts an
decisive role, complementary to inflation-targeting-oriented interest rate
feedback rules of the Taylor-type, in ensuring macroeconomic stability.
Sustainable bond-financed fiscal boosts, decreasing taxes and/or increasing
public transfers in reaction to deflationary pressures under the respect of
the public solvency condition, are capable of eradicating liquidity traps
and, at the same time, drive inflation into a globally stable trajectory
converging to the target rate. The reason is twofold. On the one hand, under
a fiscal policy regime preventing the applications of fiscal austerity
measures following deflationary slumps, a redistribution of real financial
wealth from future to current generations occurs, since deflation increases
the real value of government liabilities, thereby enhancing the burden of
future fiscal retrenchment. On the other hand, fiscal expansions financed by
bond issues and future taxes net of transfers enlarge both human and
non-human wealth for living generations. The two effects induce currently
alive individuals to dissave, spurring aggregate consumption despite the
increases in real interest rates in the vicinity of the trapping
equilibrium. Following the resulting excess demand in the goods' market,
price increases are necessary to absorb the disequilibrium.

Third, global determinacy is not guaranteed by all types of expansionary
budgetary policies that seek to fight liquidity traps. Overly aggressive
fiscal stimulus, in particular, induces spiral dynamics around the
liquidity-trap steady state. In this event, although liquidity traps are
escaped along a saddle connection from the steady state at which monetary
policy is passive to the steady state at which monetary policy is active and
convergence of inflation to the target rate is ensured, global indeterminacy
applies, thus warning against the enforcement of excessively pronounced
fiscal boosts expected to be paid back by large future primary surpluses.

Should the latter case be avoided, however, the foregoing results provide
sound analytical microfoundations to the view that demand side oriented
fiscal intervention respecting the government's intertemporal budget
constraint, because of the existence of a binding lower bound on nominal
interest rates that severely limit the power of the central bank,
constitutes a tool of `last resort' for liquidity traps to be aptly averted
and, at the same time, for aggregate stability to be preserved.

\section*{Appendix A}

Using the definition of total consumption (\ref{tc}) and the optimal
intratemporal condition (\ref{4}), the instantaneous utility function takes
the form
\begin{equation}
\log \Upsilon \left( \overline{c}(s,t),\overline{m}(s,t)\right) =\log
q(t)+\log \overline{x}(s,t),  \label{A.1}
\end{equation}%
where $q(t)\equiv \Upsilon \left( \frac{\Omega (R(t))}{\Omega (R(t))+R(t)},%
\frac{1}{\Omega (R(t))+R(t)}\right) $ is identical across all generations
and can be interpreted as the utility-based cost of living index of the
basket of physical goods and real balances. Thus, the intertemporal
optimization problem can be expressed as follows:
\begin{equation}
\underset{\{ \overline{x}(s,t)\}}{\max }\int_{0}^{\infty }\log q(t)+\log
\overline{x}(s,t)e^{-\left( \mu +\rho \right) t}dt,  \label{A.2}
\end{equation}%
subject to%
\begin{equation}
\dot{\overline{a}}(s,t)=\left( R(t)-\pi (t)+\mu \right) \overline{a}(s,t)+%
\overline{y}(s,t)-\overline{\tau }(s,t)-\overline{x}(s,t),  \label{A3}
\end{equation}%
the no-Ponzi game condition (\ref{3}), and given $\overline{a}(s,0)$. Hence,
optimality implies the Euler equation in terms of total consumption (\ref{4a}%
) and the transversality condition (\ref{4c}). Integrating forward (\ref{A3}%
) and employing both the transversality condition (\ref{4c}) and the law of
motion of total consumption (\ref{4a}) yield the level total consumption
expressed as a linear function of total wealth, that is, equation (\ref{4d}%
). From (\ref{4}), I have
\begin{equation}
\overline{x}(s,t)=\Lambda (R(t))\overline{c}(s,t),  \label{A4}
\end{equation}%
where $\Lambda (R(t))\equiv 1+R(t)/\Omega \left( R(t)\right) $.
Time-differentiating (\ref{A4}) gives%
\begin{equation}
\dot{\overline{x}}(s,t)=\Lambda ^{\prime }(R(t))\overline{c}(s,t)\dot{R}%
(t)+\Lambda (R(t))\dot{\overline{c}}(s,t).  \label{A5}
\end{equation}%
Substituting (\ref{A4}) and (\ref{A5}) into (\ref{4a}) results in the law of
motion for individual consumption (\ref{5}).

\section*{Appendix B}

Aggregate wealth in per capita terms is, by definition, given by
\begin{equation}
\overline{a}(t)=\beta \int_{-\infty }^{t}\overline{a}(s,t)e^{\beta \left(
s-t\right) }ds.  \label{B1}
\end{equation}%
Differentiating with respect to time gives
\begin{equation}
\dot{\overline{a}}(t)=\beta \overline{a}(t,t)-\beta \overline{a}(t)+\beta
\int_{-\infty }^{t}\dot{\overline{a}}(s,t)e^{\beta \left( s-t\right) }ds,
\label{B2}
\end{equation}%
Since $\overline{a}(t,t)$ is equal to zero, by assumption, using (\ref{2})
into (\ref{B2}) yields
\begin{eqnarray}
\dot{\overline{a}}(t) &=&-\beta \overline{a}(t)+\mu \overline{a}(t)+\left(
R(t)-\pi (t)\right) \overline{a}(t)+\overline{y}(t)-\overline{\tau }(t)-%
\overline{c}(t)-R(t)\overline{m}(t)  \notag \\
&=&\left( R(t)-\pi (t)-n\right) \overline{a}(t)+\overline{y}(t)-\overline{%
\tau }(t)-\overline{c}(t)-R(t)\overline{m}(t).  \label{B.3}
\end{eqnarray}%
From (\ref{4e}), the per capita aggregate consumption is given by%
\begin{equation}
\overline{c}(t)=\frac{(\mu +\rho )}{\Lambda (R(t))}\left( \overline{a}(t)+%
\overline{h}(t)\right) ,  \label{B.4}
\end{equation}%
Differentiating with respect to time the definition of per capita aggregate
consumption gives
\begin{equation}
\dot{\overline{c}}(t)=\beta \overline{c}(t,t)-\beta \overline{c}(t)+\beta
\int_{-\infty }^{t}\dot{\overline{c}}(s,t)e^{\beta \left( s-t\right) }ds,
\label{B.5}
\end{equation}%
where $\overline{c}(t,t)$ represents consumption of the newborn generation.
Because $\overline{a}(t,t)=0$ and $\overline{h}(t,t)=\overline{h}(t)$, (\ref%
{4e}) implies
\begin{equation}
\overline{c}(t,t)=\frac{(\mu +\rho )}{\Lambda (R(t))}\overline{h}(t).
\label{B.6}
\end{equation}%
Substituting (\ref{5}), (\ref{B.4}) and (\ref{B.6}) into (\ref{B.5}) yields
the time path of per capita aggregate consumption, given by (\ref{7}).

\section*{Appendix C}

Suppose that the subutility function $\Upsilon \left( \overline{c}(s,t),%
\overline{m}(s,t)\right) $ is of the CES-type, in line with Gal\'{\i} (2015)
and Walsh (2017):%
\begin{equation}
\Upsilon \left( \overline{c}(s,t),\overline{m}(s,t)\right) =\left[ \delta
\overline{c}(s,t)^{\frac{\varepsilon -1}{\varepsilon }}+\left( 1-\delta
\right) \overline{m}(s,t)^{\frac{\varepsilon -1}{\varepsilon }}\right] ^{%
\frac{\varepsilon }{\varepsilon -1}},  \label{CES}
\end{equation}%
with $0<\delta ,\varepsilon <1$, where $\varepsilon $ represents the
elasticity of substitution between consumption and real money holdings. It
then follows%
\begin{eqnarray}
\frac{\overline{m}(s,t)}{\overline{c}(s,t)} &=&\frac{1}{\Omega (R\left(
t\right) )}  \notag \\
&=&\left( \frac{\delta }{1-\delta }\right) ^{-\varepsilon }R\left( t\right)
^{-\varepsilon },  \label{money}
\end{eqnarray}%
\begin{equation}
\Lambda \left( R\left( t\right) \right) =1+\left( \frac{\delta }{1-\delta }%
\right) ^{-\varepsilon }R\left( t\right) ^{1-\varepsilon },  \label{L-}
\end{equation}%
\begin{eqnarray}
\Lambda ^{\prime }(R\left( t\right) ) &=&\left( 1-\varepsilon \right) \left(
\frac{\delta }{1-\delta }\right) ^{-\varepsilon }R\left( t\right)
^{-\varepsilon }  \notag \\
&=&\left( 1-\varepsilon \right) \frac{\overline{m}(s,t)}{\overline{c}(s,t)}.
\label{L'-}
\end{eqnarray}%
For advanced economies' annual data, in the monetary policy literature it is
common to set $\rho =0.04$ and $\Psi (\pi ^{\ast })=0.06$, (see, e.g.,
Woodford, 2003, and Benhabib, Schmitt-Groh\'{e} and Uribe, 2001). I
initially set $\varepsilon =0.5$, implying a log-log interest elasticity of
money demand of $-0.5$, consistently with the seminal paper by Lucas (2000).
Using (\ref{money}), the parameter $\delta $ is initially chosen\ so that
the implied annual consumption velocity of money is equal to unity, which is
plausible value when a broad monetary aggregate such as $M3$ is employed. So
I obtain $\delta =0.9434$. In the baseline calibration, I set $\Psi (\pi
^{L})=0.001$ and $\Psi ^{\prime }(\pi ^{L})=0.1$ and $a^{\ast }=0.6$. In
line with United Nations World Population Prospects 2019 for 2015-2020 with
reference to high-income countries,\footnote{%
https://population.un.org/wpp/.} I set $n=0.0047$ and $\mu =0.012366$,
implying a life expectancy at birth of $80.87$ years and a birth rate of $%
0.017066$. Therefore, I have%
\begin{eqnarray}
J_{22}^{\left( a^{\ast },\pi ^{L}\right) } &=&\frac{\left( \Psi ^{\prime
}(\pi ^{L})-1\right) \left[ 1+\left( \frac{\delta }{1-\delta }\right)
^{-\varepsilon }\Psi (\pi ^{L})^{1-\varepsilon }\right] }{\Psi ^{\prime
}(\pi ^{L})\left[ \left( 1-\varepsilon \right) \left( \frac{\delta }{%
1-\delta }\right) ^{-\varepsilon }\Psi (\pi ^{L})^{-\varepsilon }\right] }+%
\frac{\beta (\rho +\mu )}{1+\left( \frac{\delta }{1-\delta }\right)
^{-\varepsilon }\Psi (\pi ^{L})^{1-\varepsilon }}a^{\ast } \\
&=&-1.951.  \notag
\end{eqnarray}%
The result $J_{22}^{\left( a^{\ast },\pi ^{L}\right) }<0$ is robust to large
variations in parameter values. In particular, $J_{22}^{\left( a^{\ast },\pi
^{L}\right) }$ continues to be negative if: (a) $\varepsilon $, the
elasticity of substitution between consumption and real money holdings, is
lowered from the baseline value of $0.5$ to a value in the range $\left(
0.05,0.5\right) $, implying a diminished log-log interest elasticity of
money demand consistently with the empirical results obtained by Ireland
(2009); (b) the annual consumption velocity of money is increased from $1$
to a value in the range $\left( 1,20\right) $, should $M0$, $M1$, or $M2$ be
used as monetary aggregates; (c) $\Psi (\pi ^{L})$, the nominal interest
rate at the liquidity trap steady-state, is increased from $0.001$ to a
value in the range $\left( 0.001,0.1\right) $; (d) $\Psi ^{\prime }(\pi ^{L})
$, the monetary policy feedback reaction to inflation at the liquidity-trap
steady-state, is decreased from $0.1$ to a value in the range $\left(
0.1,0.01\right) $; (e) $a^{\ast }$, the target level for the government
liabilities to output ratio, is increased from $0.6$ to a value in the range
$\left( 0.6,2\right) $.

\section*{References}

\begin{description}
\item[\hspace{-1cm}] Barro, R. J. (1974), \textquotedblleft Are Government
Bonds Net Wealth?\textquotedblright , \textit{Journal of Political Economy}
82, 1095-1117.

\item[\hspace{-1cm}] Bartsch, E., J. Boivin, S. Fischer and P. Hildebrand
(2019), \textquotedblleft Dealing with the Next Downturn: From
Unconventional Monetary Policy to Unprecedented Policy
Coordination\textquotedblright , \textit{SUERF Policy Note} 105.

\item[\hspace{-1cm}] Bayer, C., B. Born, R. Luetticke and G. M\"{u}ller
(2020), \textquotedblleft The Coronavirus Stimulus Package: How Large is the
Transfer Multiplier?\textquotedblright , \textit{CEPR Discussion Papers}
14600.

\item[\hspace{-1cm}] {Benhabib, J., S. Schmitt-Groh\'{e}\ and M. Uribe
(2001), \textquotedblleft The Perils of Taylor Rules\textquotedblright ,
\textit{Journal of Economic Theory}} 96, 40-69.

\item[\hspace{-1cm}] {Benhabib, J., S. Schmitt-Groh\'{e}\ and M. Uribe
(2002), \textquotedblleft Avoiding Liquidity Traps\textquotedblright ,
\textit{Journal of Political Economy }}110, 535-563.

\item[\hspace{-1cm}] Bilbiie, F. (2018), \textquotedblleft Neo-Fisherian
Policies and Liquidity Traps\textquotedblright , \textit{CEPR Discussion
Papers} 13334.

\item[\hspace{-1cm}] {Blanchard, O. J. (1985), \textquotedblleft Debt,
Deficits, and Finite Horizons\textquotedblright , \textit{Journal of
Political Economy} 93, 223-247.}

\item[\hspace{-1cm}] Brock, W. A. (1974), \textquotedblleft Money and
Growth: The Case of Long-Run Perfect Foresight\textquotedblright , \textit{%
International Economic Review} 15, 750-777.

\item[\hspace{-1cm}] Brock, W. A. (1975), \textquotedblleft A Simple Perfect
Foresight Monetary Model\textquotedblright , \textit{Journal of Monetary
Economics} 1, 133-150.

\item[\hspace{-1cm}] Buiter, W. H. (2020), \textquotedblleft Three Strikes
Against the Fed\textquotedblright , VoxEU.org, 3 April.

\item[\hspace{-1cm}] Canzoneri, M., R. Cumby and B. Diba (2010),
\textquotedblleft The Interaction Between Monetary and Fiscal
Policy\textquotedblright , in B. M. Friedman and M. Woodford (Eds.), \textit{%
Handbook of Monetary Economics} 3, Amsterdam/Boston: North-Holland/Elsevier,
935-999.

\item[\hspace{-1cm}] Cheron, A., K. Nishimura, C. Nourry, T. Seegmuller and
A. Venditti (2019), \textquotedblleft Growth and Public Debt: What Are the
Relevant Trade-Offs?\textquotedblright , \textit{Journal of Money, Credit
and Banking} 51, 655-682.

\item[\hspace{-1cm}] {Cushing, M. J. (1999), \textquotedblleft The
Indeterminacy of Prices under Interest Rate Pegging: The Non-Ricardian
Case\textquotedblright , \textit{Journal of Monetary Economics} 44, 131-148.}

\item[\hspace{-1cm}] {Deaton, A. and J. Muellbauer (1980), \textit{Economics
and Consumer Behavior}, Oxford: Oxford University Press.}

\item[\hspace{-1cm}] {Gal\'{\i}, J. (2015), \textit{Monetary Policy,
Inflation and the Business Cycle}, Princeton: Princeton University Press.}

\item[\hspace{-1cm}] Ireland, P. N. (2009), \textquotedblleft On the Welfare
Cost of Inflation and the Recent Behavior of Money Demand\textquotedblright
, \textit{American Economic Review} 99, 1040-1052.

\item[\hspace{-1cm}] Krugman, P. (2020), \textquotedblleft The Case for
Permanent Stimulus\textquotedblright , in R. Baldwin and B. Weder di Mauro
(Eds.), \textit{Mitigating the COVID Economic Crisis: Act Fast and Do
Whatever It Takes} 3, London: CEPR Press, 213-219.

\item[\hspace{-1cm}] Leeper, E. M. (1991), \textquotedblleft Equilibria
under 'Active' and 'Passive' Monetary and Fiscal Policies\textquotedblright
, \textit{Journal of Monetary Economics} 27, 129-147.

\item[\hspace{-1cm}] Lucas, R. E. (2000), \textquotedblleft Inflation and
Welfare\textquotedblright , \textit{Econometrica} 68, 247-274.

\item[\hspace{-1cm}] Maebayashi, N., T. Hori and K. Futagami (2017),
\textquotedblleft Dynamic Analysis of Reductions in Public Debt in an
Endogenous Growth Model with Public Capital\textquotedblright , \textit{%
Macroeconomic Dynamics} 21, 1454-1483.

\item[\hspace{-1cm}] {Marini, G. and F. Van der Ploeg (1988),
\textquotedblleft Monetary and Fiscal Policy in an Optimizing Model with
Capital Accumulation and Finite Lives\textquotedblright , \textit{Economic
Journal} 98, 772-786.}

\item[\hspace{-1cm}] Mertens, K. and Ravn, M. (2014), \textquotedblleft
Fiscal Policy in an Expectations-Driven Liquidity Trap\textquotedblright ,
\textit{Review of Economic Studies} 81, 1637-1667.

\item[\hspace{-1cm}] Minea, A. and P. Villieu (2013), \textquotedblleft Debt
Policy Rule, Productive Government Spending, and Multiple Growth Paths: A
Note\textquotedblright , \textit{Macroeconomic Dynamics} 17, 947-954.

\item[\hspace{-1cm}] Nakata, T. and Schmidt, S. (2019), \textquotedblleft
Expectations-Driven Liquidity Traps: Implications for Monetary and Fiscal
Policy\textquotedblright , \textit{ECB Working Paper Series} 2304.

\item[\hspace{-1cm}] Patinkin, D. (1965), \textit{Money, Interest, and Prices%
}, New York: Harper and Row.

\item[\hspace{-1cm}] Pigou, A. C. (1943), \textquotedblleft The Classical
Stationary State\textquotedblright , \textit{The Economic Journal} 53,
343-351.

\item[\hspace{-1cm}] Pigou, A. C. (1947), \textquotedblleft Economic
Progress in a Stable Environment\textquotedblright ,\  \textit{Economica} 14,
180-188.

\item[\hspace{-1cm}] {Reis, R. (2007), \textquotedblleft The Analytics of
Monetary Non-Neutrality in the Sidrauski Model\textquotedblright , \textit{%
Economics Letters} 94, 129-135.}

\item[\hspace{-1cm}] Sargent, T. J. and N. Wallace (1981), \textquotedblleft
Some Unpleasant Monetarist Arithmetic\textquotedblright , Federal Reserve
Bank of Minneapolis \textit{Quarterly Review} 5, 1-17.

\item[\hspace{-1cm}] Schmidt, S. (2016), \textquotedblleft Lack of
Confidence, the Zero Lower Bound, and the Virtue of Fiscal
Rules\textquotedblright , \textit{Journal of Economic Dynamics and Control}
70, 36-53.

\item[\hspace{-1cm}] Schmitt-Groh\'{e}, S. and Uribe, M. (2009),
\textquotedblleft Liquidity Traps with Global Taylor Rules\textquotedblright
, \textit{International Journal of Economic Theory}, 85-106.

\item[\hspace{-1cm}] Sidrauski, M. (1967), \textquotedblleft Rational Choice
and Patterns of Growth in a Monetary Economy\textquotedblright , \textit{%
American Economic Review} 57, 534-544.

\item[\hspace{-1cm}] Summers, L. H. (2016), \textquotedblleft Secular
Stagnation and Monetary Policy\textquotedblright , Federal Reserve Bank of
St. Louis \textit{Review} 98, 93-110.

\item[\hspace{-1cm}] Summers, L. H. (2018), \textquotedblleft Secular
Stagnation and Macroeconomic Policy\textquotedblright , \textit{IMF Economic
Review} 66, 226-250.

\item[\hspace{-1cm}] {Taylor}, J. B. (Ed.) (1999), \textit{Monetary Policy
Rules}, Chicago: University of Chicago Press.

\item[\hspace{-1cm}] {Taylor, J. B. (2012), \textquotedblleft Monetary
Policy Rules Work and Discretion Doesn't: A Tale of Two
Eras\textquotedblright , \textit{Journal of Money Credit and Banking} 44,
1017-1032.}

\item[\hspace{-1cm}] Taylor, J. B. (2018), \textquotedblleft Fiscal Stimulus
Programs During the Great Recession\textquotedblright , \textit{Economics
Working Paper} 18117, Hoover Institution.

\item[\hspace{-1cm}] Taylor, J. B. (2021), \textquotedblleft Simple Monetary
Rules: Many Strengths and Few Weakenesses\textquotedblright , \textit{%
European Journal of Law and Economics}, forthcoming.

\item[\hspace{-1cm}] Walsh, C. E. (2017), \textit{Monetary Theory and Policy}%
, Cambridge: The MIT Press.

\item[\hspace{-1cm}] {Weil, P. (1989), \textquotedblleft Overlapping
Families of Infinitely-Lived Agents\textquotedblright , \textit{Journal of
Public Economics} 38, 183-198.}

\item[\hspace{-1cm}] Woodford, M. (2003), \textit{Interest and Prices},
Princeton: Princeton University Press.

\item[\hspace{-1cm}] {Yaari, M. E. (1965), \textquotedblleft Uncertain
Lifetime, Life Insurance, and the Theory of the Consumer\textquotedblright ,
\textit{The Review of Economic Studies} 32, 137-150.}
\end{description}

%\end{doublespace}

\end{document}